%% file: wb86r.arxiv.tex
\documentclass[preprint,12pt]{aastex62}		         

\input{epsf}
 
\usepackage{natbib}
\usepackage{graphicx}	
\usepackage{amsmath}

\input{apj_commands.tex}
\shorttitle{Orientation effects in quasars}
\shortauthors{Runnoe et al.}

\begin{document}		
\title{Orientation and accretion in a representative sample of active galactic nuclei}

\correspondingauthor{Jessie C. Runnoe}
\author{Jessie C. Runnoe}
\affil{Department of Physics and Astronomy, Vanderbilt University, Nashville, TN 37235}
\email{jessie.c.runnoe@vanderbilt.edu}

\author{Todd Boroson}
\affil{Las Cumbres Observatory Global Telescope Network, Goleta, CA 93117}


\begin{abstract}
We highlight a representative sample of active galactic nuclei selected independent of orientation.  The defining characteristic of the selection is sophisticated matching between the $0.1<z<0.6$ Sloan Digital Sky Survey quasars from the Seventh Data Release to the Westerbork Northern Sky Survey at 325 MHz and the subsequent application of a total radio luminosity cut.  The resulting sample is complete down to the limiting luminosity and unbiased by orientation.  Compared to orientation samples in the literature this approach yields less bias with redshift, relatively more lobe-dominated sources including those with radio lobes and no visible core, and a distribution of radio core dominance that is consistent with expectations from a uniform distribution of inclinations with solid angle.  We measure properties of the optical spectra, and use the sample to investigate the orientation dependence of the velocity width of the broad \Hb\ emission line.  We recover the known orientation dependence, but the sharp envelope of previous studies where only edge-on sources display the broadest lines, is absent.  Scatter in this diagram is not attributable solely to black hole mass, Eddington ratio, or contamination in the sample from compact steep spectrum sources.  A physical framework for quasar beaming and a disk-like broad-line region can describe the representative sample when it is expanded to include additional parameters, in particular jet properties and the broad-line region velocity field.  These points serve to illustrate the critical role of sample selection in the interpretation of observable properties as indicators of physical parameters of quasar central engines.
\end{abstract}

\section{Introduction}
For active galactic nuclei (AGN), unification by orientation implies that all AGN are intrinsically the same and apparent differences are the result of different viewing angles.  In this spirit, orientation has been invoked to explain the variety of AGN classes.  For radio-loud (RL) sources, core-dominated quasars, lobe-dominated quasars, and radio galaxies have been thought to represent a continuum in inclination angle from jet-on to edge-on \citep[although see][]{dipompeo13d}.  Similarly, \citet{antonucci85} uncovered the first hidden broad-line region (BLR) in NGC~1068, a Type 2 source, demonstrating that the optical Type 1 and Type 2 spectral classifications can result from the difference between a face-on and edge-on view where  central engine is obscured \citep[although see also][and references therein for cases where the Type 1/Type 2 dichotomy is not due to orientation]{runnoe16a}.  On the other hand, there is known variance among quasar broad and narrow emission lines, which originate in the BLR and narrow-line region (NLR), respectively, that cannot be ascribed to orientation effects \citep{bg92}.  Likely the truth lies somewhere between the two extremes: orientation plays a role in determining the observable properties of quasars and AGN, but is not the only important factor.  The difficulty in teasing out the relative contributions from different mechanisms is in selecting an appropriate sample where orientation and other properties can all be measured.  

The viewing angles in the approximately 10\% of quasars that are RL can be estimated from the radio spectrum by taking advantage of anisotropic relativistic beaming effects \citep[e.g.,][and references therein]{orr82,wills95,vangorkom15}.  Radio core dominance and radio spectral index, the two most common radio orientation indicators, both estimate orientation by quantifying the amount of Doppler boosting in the radio core that increases with more jet-on lines of sight.  Radio core dominance does this simply by measuring the amount of beaming in the core normalized to the intrinsic luminosity of the source \citep{orr82,wills95}.  The radio spectral index measurement characterizes the shape of the radio spectrum; more jet-on orientations have flat radio spectra, the result of boosting the superimposed synchrotron self-absorbed spectra associated with the radio core, while more edge-on orientations show the steep radio spectra associated with the isotropic optically-thin emission from the radio lobes.  In both cases, the measurement is statistical in nature and the correlation with viewing angle includes scatter \citep{willsbro95,fine11,dipompeo12a,marin16}. 

The combination of radio observations and optical spectroscopy in a well-defined sample yields considerable insight into the geometry of the BLR.  It has been shown that the broad permitted emission lines (namely but not exclusively \Hb) are broadest and have more complex profiles in lobe-dominated sources \citep{miley79,boroson84,boroson85}.  In a seminal paper, \citet[][hereafter WB86]{wills86} measured radio core dominance in a sample of quasars with rest-frame optical spectra covering \Hb\ and found that there is an envelope in radio core dominance versus full-width at half maximum (FWHM) space, such that jet-dominated sources have relatively narrow broad lines, while lobe-dominated sources have a range of line widths.  Based on this evidence, those authors infered a geometry where the broad-line emitting gas is confined to a flattened disk with its axis aligned with the radio jet.  Subsequently, this dependence was reproduced in other samples \citep[e.g.,][]{runnoe13a}, including those with likely applications to radio-quiet (RQ) quasars \citep{jarvis06,fine11}, and for other broad emission lines \citep{vestergaard00,runnoe14}.

However, orientation is not the only mechanism that determines the observable properties of quasar spectra, which are very diverse \citep{bg92,robinson95,sulentic03,shen14}.  Using principal component analysis, \citet{bg92} characterized the object-to-object variation observed in quasar spectra.  The first component from this analysis, Eigenvector 1 (EV1), is most apparent as an anti-correlation between optical \FeII\ and \OIIIw\ and has since been connected to a host of additional properties in other wavebands \citep[e.g.,][]{brandt00}.  Although the physical mechanism(s) underlying EV1 are not completely understood, the popular interpretation is that Eddington ratio ($L_{bol}/L_{Edd}$), which is strongly correlated with EV1, governs the vertical structure of the accretion disk and thus the illumination of the external disk, BLR, and NLR.  While EV1 itself is not driven exclusively by orientation \citep{bg92,boroson92}, neither does it appear to be completely orientation independent.  The spectral proxies commonly used to describe EV1, for example $R_{\textrm{\FeII}} \equiv \textrm{EW}(\textrm{\FeII}/\textrm{\Hb})$, do have subtle orientation dependencies \citep{runnoe14} likely due to anisotropy in the \FeII\ emission \citep{joly91} rather than the continuum which drops out of the equivalent width (EW) ratio.  As a result, the relative contributions of orientation and other drivers of EV1 in determining the optical spectra of quasars are not completely clear.
	
Investigations seeking to examine these effects are complicated by the difficulty of selecting an appropriate sample.  Solutions to this problem have varied, often driven by current observational capabilities.  Early samples tended to include all possible objects because it was initially very difficult to collect the requisite optical and radio observations.  Technical limitations resulted in a tendency to include objects with good optical identifications, typically either objects from the smallest flux-limited surveys or those with well-known positions which favored point-like radio sources.  

With the advent of large-area optical and radio surveys, samples are now often constructed by matching the Sloan Digital Sky Survey \citep[SDSS;][]{york00} with Faint Images of the Radio Sky at Twenty-cm \citep[FIRST;][]{becker95} in some radius, often $2-3$\arcsec\ \citep[e.g.,][]{brotherton15}.  Sometimes an additional cut is imposed in order to ensure the availability of low-frequency radio observations from, for example, the Westerbork Northern Sky Survey \citep[WENSS,][]{rengelink97} at 325~MHz \citep[e.g.,][]{jarvis06,fine11}.  These samples are complete to some flux limit over a large portion of the sky which introduces a selection bias: the limiting luminosity is a strong function of redshift and at higher redshifts lobe-dominated objects will preferentially fall below the detection limit \citep{lu07}.  The result of this selection method is that a substantial number of objects with lobes and no core may not be identified by the matching process \citep{lu07,jackson13} and at higher redshifts core-dominated objects will be preferentially selected at a given luminosity (particularly when selected at higher radio frequencies).  More sophisticated matching algorithms can ameliorate these issues \citep{devries06,lu07,kimball11}, but when conducted at high radio frequency may still be biased against lobe-dominated objects because of their steep spectra or because the small beam size of FIRST is not suited to detecting extended emission.  

A third approach is to study orientation effects by selecting intrinsically similar objects, based on their extended radio luminosity, so that changes in the radio core dominance parameter are the result of different amounts of beaming \citep{wills95,netzer95,shang11,runnoe13a,runnoe13d,runnoe14}. The strength of this approach is that it should isolate orientation effects.  Such a sample is likely a fair representation of the more edge-on sources, but will make the jet-on sources seem more common than they actually are in an effort to obtain a range in the orientation parameter. Furthermore, when selected from a flux limited sample, these samples may inherit the redshift-luminosity dependence.

The goal of this work is to select a sample of sources in a way that minimizes the biases described above and investigate the effect of orientation on their optical spectra.  This paper is organized as follows.  The sample selection, spectral decomposition, and measurements are described in Section~\ref{sec:sample}.  In Section~\ref{sec:compsamp} we outline several comparison samples from the literature and detail extra measurements that we compile for them.  We discuss the unique properties of the sample in the context of comparison samples in Section~\ref{sec:prop} and present the orientation dependencies of various spectral properties in Section~\ref{sec:analysis}.  Section~\ref{sec:wb86r} describes the updates to the parameters of beaming models from the literature that are required to describe the representative sample in this context.  We discuss the advantages and limitations of the analysis in the context of previous work in Section~\ref{sec:discuss} and summarize our results in Section~\ref{sec:summary}.  Throughout this work, we adopt a cosmology of with $H_0 = 71$ km s$^{-1}$ Mpc$^{-1}$, $\Omega_{\Lambda} = 0.73$, and $\Omega_{m} = 0.27$.

\section{A representative orientation sample}
\label{sec:sample}
\subsection{Sample selection}
We selected objects from the seventh data release (DR7) quasar catalog \citep{schneider07} of the Sloan Digital Sky Survey \citep[SDSS,][]{york00}.  We first applied a redshift cut of $0.1<z<0.6$ to ensure complete spectral coverage of the \Hb\ region for all sources.  

To target emission from the extended radio structures, which is thought to be emitted isotropically, we matched to WENSS at 325~MHz.  The radio structures associated with a single SDSS quasar may appear as multiple entries in the WENSS catalog, so the match was done in two steps of identifying candidate cores and lobes and then confirming them.  Core candidates were matched within 30\arcsec\ and lobe candidates were identified by matching within 1100\arcsec, an angular distance that corresponds to 2.9~Mpc at the low-redshift limit of the sample, and requiring a total size less than 3~Mpc.  The large match radii are meant to be inclusive; radio sources larger than 3~Mpc are very uncommon \citep{wardle74,riley89,kuzmicz18,doi19} so all relevant WENSS catalog entires will be identified with spurious associations to be pruned later.  When two lobe candidates were associated with one core, we also required that they be within 30$^\circ$ of being opposite each other and have fluxes and distances from the SDSS position within a factor of two.  A luminosity cut of log$(L_{325})<33.0$ erg~s${-1}$~Hz$^{-1}$, where $L_{325}$ is the total radio luminosity at 325~MHz, keeps the sample well above the WENSS flux limit of 18~mJy.  We note that this luminosity is also well above the traditional transition between FR\;I and FR\;II sources at log$(L_{325})\sim32.5$, where intrinsic differences might be expected \citep{fanaroff74}.  There were 142 sources that matched these criteria.  

Each source was then visually inspected in the SDSS, WENSS, FIRST, and the NRAO VLA Sky Survey \citep[NVSS,][]{condon98} to verify that all components matched correctly and collect the corresponding FIRST and NVSS fluxes.  Search radii for matching to FIRST and NVSS were determined manually to ensure that all components, which are sometimes listed as separate catalog entries, were collected.  When no core component was detected in FIRST, we used the FIRST RMS and adopted a $5\sigma$ limit on the core measurement.  Finally, the luminosity cut was re-evaluated to ensure that the luminosity was not over-estimated in cases where lobe candidates were identified as a mismatch and removed.  

The final sample includes 126 objects and constitutes a complete sample in the sense that it contains all objects meeting the above criteria down to the limiting luminosity.  Properties related to the sample selection are presented in Table~\ref{tab:sample}.  

{\rotate
\input{all_prop_short.tex}
}


\subsection{Radio properties}	
\label{sec:radprop}
In addition to listing objects selected to the sample, Table~\ref{tab:sample} also includes several calculated radio properties for each source.  For these calculations, we adopt the convention $S_{\nu}\propto\nu^{\alpha_r}$ for the radio spectrum and apply k corrections by adjusting eqn.~1 of \citet{stocke92} to the relevant observed and desired frequencies.  The low-frequency total radio luminosity density, $L_{325}$, at 325~MHz rest frame, is a measure of the total luminosity of the source, presumably unaffected by beaming of the core which will not contribute significantly at such low frequencies.  This quantity was calculated from the total WENSS flux density assuming a spectral slope of $\alpha_r=-1$.  The core luminosity density, $L_{c}$, at 5~GHz rest frame, is a measure of the beamed core luminosity.  This quantity was determined from the FIRST core component (i.e. the FIRST peak flux of the component visually determined to be the core) and k-corrected to 5~GHz rest frame assuming a flat spectrum.  When no core component was detected, we adopted five times the RMS value of the nearest component as a limit on the core flux density.

Also listed in Table~\ref{tab:sample} are two orientation indicators, radio core dominance and radio spectral index.  These both rely on the fact that relativistic beaming in the radio jet is an anisotropic process, with beaming increasing at more jet-on lines of sight.  Radio core dominance is an estimate of the amount of beaming present at a given intrinsic luminosity and requires a resolved source in order to make a measurement.  In its original form, the extended radio luminosity at 5~GHz rest frame was used as a normalizing factor \citep{orr82}.  However, according to \citet{willsbro95}, using a normalizing luminosity calculated in the optical band provides a superior measure of orientation because it avoids the effects of interaction between the extended radio emitting structures and the interstellar medium of the host galaxy.  Indeed, in a recent test of the efficacy of various radio orientation indicators \citet{vangorkom15} identified this as the best approach.  From \citet{willsbro95}, 

\begin{eqnarray}
\textrm{log}\left(R_V\right) = \textrm{log}\left(\frac{L_c}{\textrm{erg s}^{-1}}\right)+M_V/2.5-21.69 = \textrm{log}\left(\frac{L_c}{L_{opt}}\right),
\end{eqnarray}

\noindent where $M_{V}$ is the extinction-corrected, k-corrected absolute magnitude in the $V$ band and $L_{opt}$ is the optical continuum luminosity of the quasar.  Here, we take the monochromatic luminosity of the quasar continuum at 5100~\AA, $L_{5100}$, determined from the spectral decomposition (see Section~\ref{sec:sfit}) as $L_{opt}$.

Radio spectral index is a measure of the spectral slope in the radio, which is flatter for more jet-on sources and steeper for more edge-on sources.  It does not require a resolved source, but there is a risk of incorrectly attributing different emission components.  Adopting the convention $S_{\nu} \sim \nu^{\alpha_{r}}$, we measured radio spectral index, $\alpha_{r}$, of the total flux between 325~MHz and 1.4~GHz.

Uncertainties in the radio properties are taken to be the standard deviation of the distribution for each measurement determined by bootstrap resampling the radio or optical SDSS fluxes and re-calculating the relevant quantity $10^4$ times.  For WENSS, we took the uncertainty on the flux measurements to be 3.6~mJy, which is the level of the noise in the radio maps.

\subsection{Optical spectral decomposition and measurements}
\label{sec:sfit}
We performed a spectral decomposition so that we could quantify the properties of the optical continuum and emission lines, namely \Hb, \OIII, and \FeII.  Before decomposing the SDSS spectra and making measurements, we applied two corrections.  First, we performed a Galactic extinction correction using the dust maps of \citet{schlegel98} and assuming a Seaton reddening law \citep{seaton79}.  We then shifted the spectra to the rest frame using the SDSS redshifts, retaining the observed-frame fluxes.  

The spectral decomposition was done following the methodology of \citet{runnoe15a}, which is particularly effective at separating fit components in objects where the broad lines are weak or complex.  In this approach, the fitting uses the $\chi^2$ minimization IRAF task \textsc{specfit} \citep{kriss94} and is performed in three steps.

In the first step, we fit and subtracted the optical continuum, which includes contributions from the AGN continuum, optical \FeII, and stars in the host galaxy.  The model consisted of a featureless power-law continuum, the optical \FeII\ template of \citet{veron-cetty04}, and a Bruzual \& Charlot instantaneous starburst (SB) template \citep[e.g.,][]{cales12} where the age and mass are variable parameters.  The inclusion of the SB template was our only departure from the \citet{runnoe15a} procedure and was made only when the addition of free parameters was warranted based on the Bayesian Information Criterion \citep{schwartz78}.  Continuum-subtracted spectra were generated by subtracting the best-fit optical continuum model from the data.

In the second step, we characterized the \OIIIdblt\ emission and subtracted it.  We parameterized the profile of narrow emission in the spectrum based on the \OIIIw\ emission line. In order to isolate that line, a low-order polynomial was fit to the local continuum (i.e. the red wing of \Hb) in manually selected wavelength windows.  The polynomial was subtracted and the remaining emission line was fit with two Gaussians, to which no physical meaning was attributed, to account for possible line asymmetry.  The \OIII\:$\lambda4959$ line profile was then taken to be identical to that of \OIIIw, but appropriately shifted in wavelength and scaled down in flux by a factor of $1/3$.  The \OIII\ doublet (but not the low-order polynomial continuum) was subtracted to generate a clean spectrum with which to model the \Hb\ emission.

In the third and final step, we decomposed the emission from \Hb.  The \Hb\ emission was fit in most objects with a combination of four Gaussian components.  Two of the Gaussians accounted for the narrow \Hb\ emission.  These were constrained to create a profile identical in shape and tied in wavelength to the \OIIIw\ line, although the flux of the profile was allowed to vary and even to be zero if it was the best fit.  The other two Gaussians were allowed to vary freely and characterize the broad emission.  In a few cases where the broad \Hb\ line profiles were very boxy, an additional Gaussian was necessary to adequately characterize the broad \Hb.

\begin{figure}[htb]
\begin{center}
\includegraphics[width=15cm]{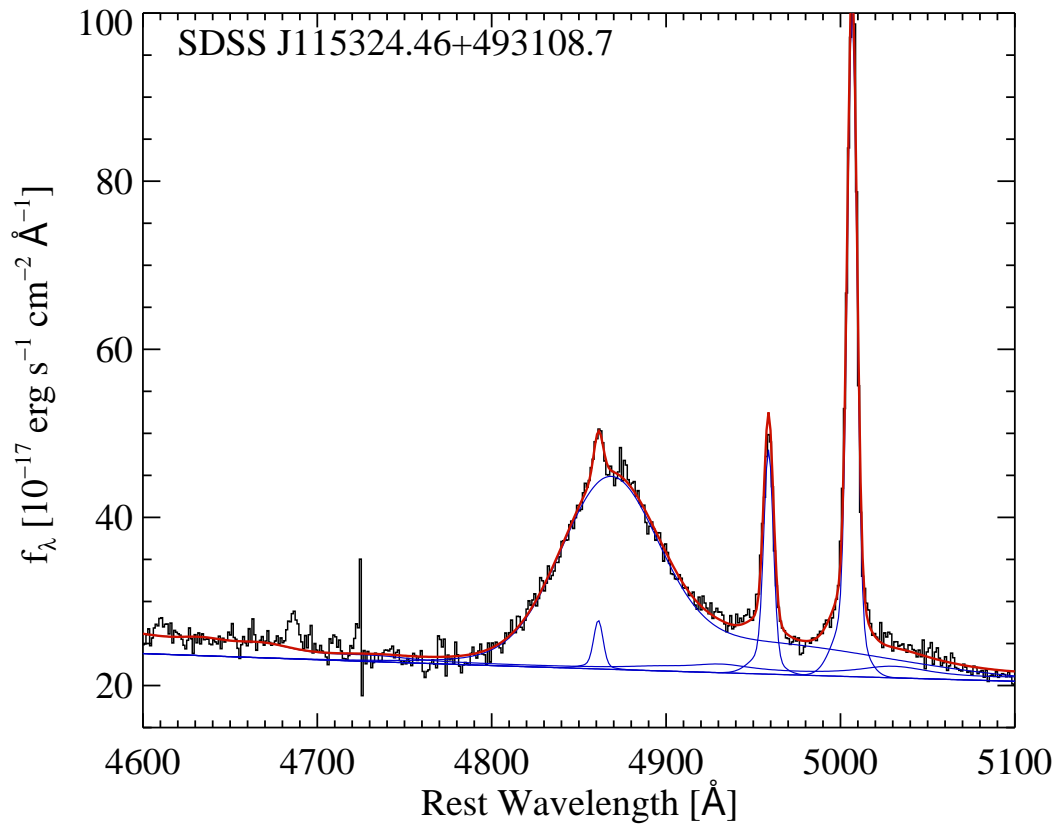}
\caption{Example of the spectral decomposition.  The red line is the total model and the blue lines show the power-law continuum, \FeII\ template, and broad and narrow emission line components.  \label{fig:specfit}}
\end{center}
\end{figure}

We visually inspected the final fits, with all components combined, to verify that the result faithfully reproduced the data.  The result of this process can be viewed in Figure~\ref{fig:specfit} for one representative object.

Following \citet{runnoe15a}, we measured spectral properties from the high S/N model spectra in a window where the model flux is above 1\% of the peak flux density for the line of interest.  When measuring the EWs, the continuum was taken to be the power-law model from the fit, and the \FeII\ EW was calculated between 4435 and 4685~\AA.  Uncertainties on the spectral properties were calculated from the S/N in the \Hb\ emission line using the scaling relationships of \citet{runnoe15a}, which were derived from a Monte-Carlo simulation of the spectral decomposition procedure, including the manual steps.  The final 5100~\AA\ continuum luminosity measurements, $\lambda L_{\lambda}$, are listed with other luminosity quantities in Table~\ref{tab:sample} and the FWHM and rest-frame EW measurements are listed in Table~\ref{tab:spec}.

\input{spec_prop_short.tex}

In some cases, the continuum decomposition appears degenerate.  That is, while we have reliably modeled the total continuum, the individual components (power law, \FeII\ template, and stellar template) may not be robustly separated.  For this reason, whenever possible we based our analysis on EW ratios for nearby emission lines, where the continuum value drops out.  In the case of the optical continuum luminosity used to determine the radio core dominance parameter, we have experimented with other characterizations of the optical luminosity (e.g., by calculating absolute magnitudes from SDSS magnitudes and average spectral shapes) , or radio core dominance parameters based solely on radio observations and find that our overall conclusions are robust against these choices.

\section{Comparison samples and data}
\label{sec:compsamp}
In order to put this new sample in context, we compare it to a variety of orientation samples from the literature.  Below we describe them and the data that we tabulated to facilitate the comparison.

\subsection{The \citet{wills86} sample}
The WB86 sample cemented the paradigm for a disk-like BLR structure by showing that the velocity width of the broad H$\beta$ line depends on radio core dominance.  There was no formal selection procedure, but sample members are quasars and radio galaxies with high-quality optical and radio fluxes and many were taken from low radio frequency surveys.  The sample includes 80 objects with $0.033<z<0.960$ and has measures of redshift, H$\beta$ FWHM, and core/extended radio fluxes at 5~GHz.  

We adopted the FWHM measurements directly and re-calculated the radio core dominance from listed radio fluxes assuming core and extended spectral indices of 0 and $-1.0$, respectively.  Because we assumed the same spectral indices as \citet{wills86}, these match published values except for small differences when tabulated fluxes and radio core dominance values did not agree.  We adopted the relation from \citet{willsbro95} of log$(R)$ = log$(R_{V})^{1.10\pm0.08}-2.7\pm0.2$ to calculate log$(R_{V})$ for this sample.  We calculated the 325~MHz luminosity using radio fluxes tabulated in the NASA/IPAC Extragalactic Database\footnote{The NASA/IPAC Extragalactic Database (NED) is operated by the Jet Propulsion Laboratory, California Institute of Technology, under contract with the National Aeronautics and Space Administration.}.  Data were primarily from the TEXAS survey of radio sources at 365~MHz \citep{douglas96}.  The TEXAS survey maps the sky at $-35.5<\delta<+71.5^{\circ}$(B1950), has positional accuracy of about 1\arcsec, spatial resolution of 22\farcs1, and is 90\% complete down to 0.4~Jy.  Other fluxes were from WENSS and the Molonglo Reference Catalogue of radio sources \citep{large81} at 408~MHz.  Fluxes from sources other than WENSS were k-corrected to 325~MHz assuming $\alpha_{r}=-1.0$.  Properties for this sample are listed in Table~\ref{tab:sample_wb}.

\input{wb86_sample_short.tex}

\subsection{Sample of intrinsically similar objects based on extended radio luminosity}
The radio-loud subsample from \citet{shang11}, which we refer to as the SED sample, includes intrinsically similar objects based on their extended radio luminosity.  It was compiled \citep{wills95,netzer95} for studying orientation effects because variation in the radio core dominance parameter is attributable almost exclusively to variation in the amount of beaming.  First presented in \citet{shang11}, it was subsequently used to study orientation effects in quasar single-epoch black hole mass estimates, UV spectroscopic properties, and spectral energy distributions \citep{runnoe13a,runnoe13d,runnoe14}.  The sample includes 52 radio-loud quasars with $0.1576<z<1.404$.  

We calculated the 325~MHz luminosity for comparison with this work using radio fluxes tabulated in \citet{shang11}.  21/52 objects in this sample are covered by WENSS, for the remaining 31 objects we adopted fluxes from the TEXAS survey of radio sources at 365~MHz \citep{douglas96} and k-corrected them to 325~MHz assuming $\alpha_{r}=-1.0$.  Compared to WENSS, the spatial resolution is a factor of 2 better, but the sensitivity is $\sim10$ times worse and poor $uv$ coverage means that irregularly shaped sidelobes can make fluxes for extended sources less reliable.  For this reason, we prioritized WENSS observations whenever possible.

The spectral decomposition, 5100~\AA\ fluxes, and H$\beta$ FWHM for this sample were presented by \citet{tang12}.  We calculated the 5100~\AA\ luminosity and radio core dominance parameter, $R_{V}$, following the procedure described in Section~\ref{sec:radprop}.  Properties for this sample are listed in Table~\ref{tab:sample_sed}.  Although we ultimately do not replot the FWHM values for this sample (they can be seen in \citealt{runnoe13a}) we include them for completeness.  

\input{sed_sample_short.tex}

\subsection{Survey sample selected at high radio frequency}
We refer to this sample, originally presented in \citet{brotherton15}, as the B15 sample.  It is just one example of the behaviors that might be expected by selecting radio orientation samples at high radio frequency, often by matching SDSS and surveys like NVSS and FIRST at 1.4~GHz.  Specifically, this sample was selected from the $z<0.75$, $i<19$~mag, optically non-reddened, DR7 radio-loud quasars with spectral measurements in \citet{shen11}.  The authors cross-matched with NVSS using a 30\arcsec\ radius and keeping objects with $S_{1.4}>10$~mJy \citep{kimball08}.  They matched to FIRST within 2\arcsec, and visually inspected all of the FIRST maps to identify catalog entries that were attributable to extended radio lobes.  This selection procedure yields 386 radio-loud quasars with $0.084 < z < 0.7471$.

\citet{brotherton15} presented the sample with measurements of H$\beta$ FWHM, radio core flux, extended radio flux, radio core dominance, and 5100~\AA\ luminosity.  The FWHM parameter with associated uncertainties we adopted directly.  We calculated the radio core dominance $R_{V}$ parameter from the peak core flux and 5100~\AA\ luminosity for our assumptions and cosmology.  We also calculated the 325~MHz luminosity for 136/386 objects by matching within 3\arcsec\ to the FIRST positions in version 2 of the unified radio catalog \citep{kimball14}.  We then adopted the corresponding WENSS flux from the catalog, which was obtained by matching within 120\arcsec.  We note that this approach to matching WENSS fluxes is less comprehensive than our other methods, which may lead to some slight differences in the 325~MHz luminosity for these sources.  We accept this, since this sample is adopted largely for illustrative purposes.  Objects with 325~MHz coverage are determined by sky position in the overlapping SDSS and WENSS region, which should produce an unbiased view of the low-frequency behavior for this sample.  Properties for this sample are listed in Table~\ref{tab:sample_b15}.  Although we ultimately do not replot the FWHM values for this sample (they can be seen in \citealt{brotherton15}) we include them for completeness.  

\input{b15_sample_short.tex}

\section{Properties of the radio sample}
\label{sec:prop}
One of our primary goals is to create a sample independent of orientation angle that has properties representative of the general population of quasars.  Here, we highlight the properties of the final sample that are specific to this selection and would not have been obtained via other approaches.

\subsection{Luminosity-redshift bias}
\label{sec:Lz}
Many orientation samples are selected from flux limited surveys.  Even samples selected in other ways, like the SED sample which is chosen to have a small range in luminosity, can inherit biases when sources are taken from flux-limited samples.

\begin{figure}[htb]
\begin{center}
\includegraphics[width=15cm]{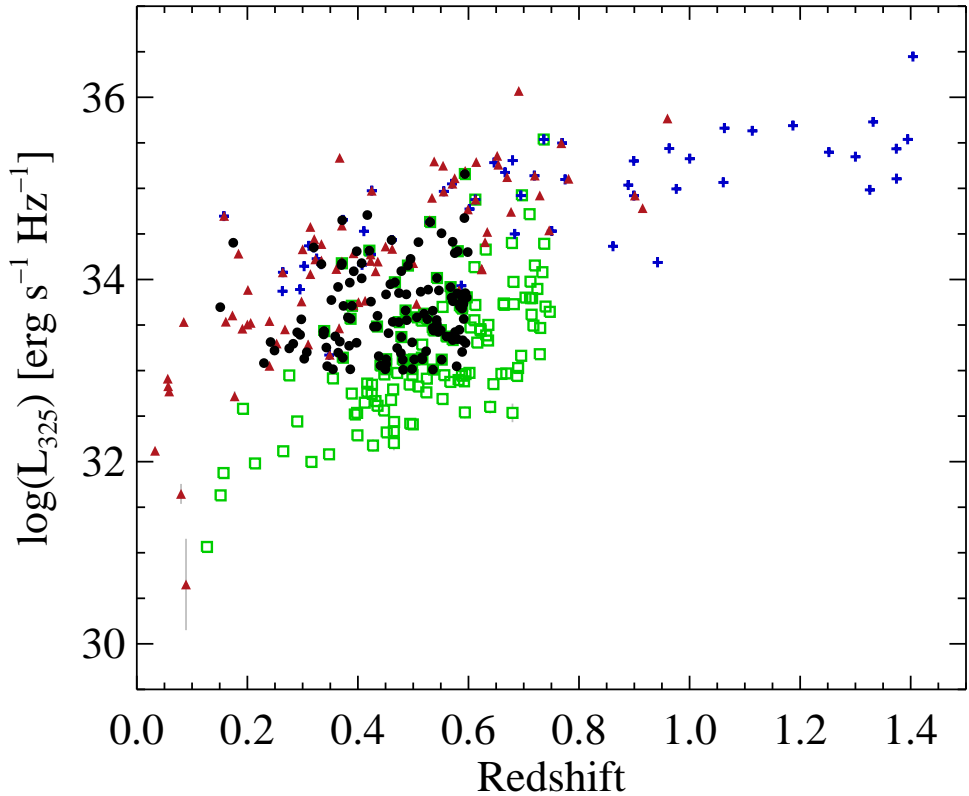}
\caption{The total low-frequency radio luminosity as a function of redshift for the new representative sample (black circles), WB86 sample (red triangles), SED sample (blue crosses), and the B15 with WENSS coverage (green squares).  Radio fluxes are well known, so uncertainties are almost always smaller than the data points.  For black circles, although the upper limit on luminosity increases with redshift as objects are chosen from a greater volume, the lower limit on the luminosity does not increase as it would in a flux-limited sample.  Also notable, the SED sample probes the highest luminosities at every redshift and has inherited some of the flux-limit bias.  \label{fig:zbias}}
\end{center}
\end{figure}

In Figure~\ref{fig:zbias} we show the total radio luminosity density at 325~MHz as a function of redshift.  For the new sample presented in this work, the maximum luminosity is a strong function of redshift because larger redshifts probe larger volumes, but the low-luminosity limit remains constant over the entire redshift range.  This is in contrast to the WB86, SED, and B15 samples where the luminosity is a strong function of the redshift as only the most luminous objects at each redshift are included in the sample.  Notably, the SED sample is very high luminosity, meaning that while the selection may help to isolate orientation effects, the behavior may differ from more common but less luminous sources.

\subsection{Biases in extended sources from flux limited selection}
\label{sec:Rz}
Flux-limited selection at high radio frequency and cross-matching within a small radius can lead to more than just luminosity biases.  In particular, a substantial number of objects with lobes and no core will be missed and at higher redshifts, core-dominated objects will be preferentially selected at a given luminosity.  That is, at high redshift lobe-dominated sources will drop out of the sample and core-dominated sources will get beamed in.  Thus, the radio core dominance parameter will also be a function of redshift.

\begin{figure}[htb]
\begin{center}
\includegraphics[width=8.1cm]{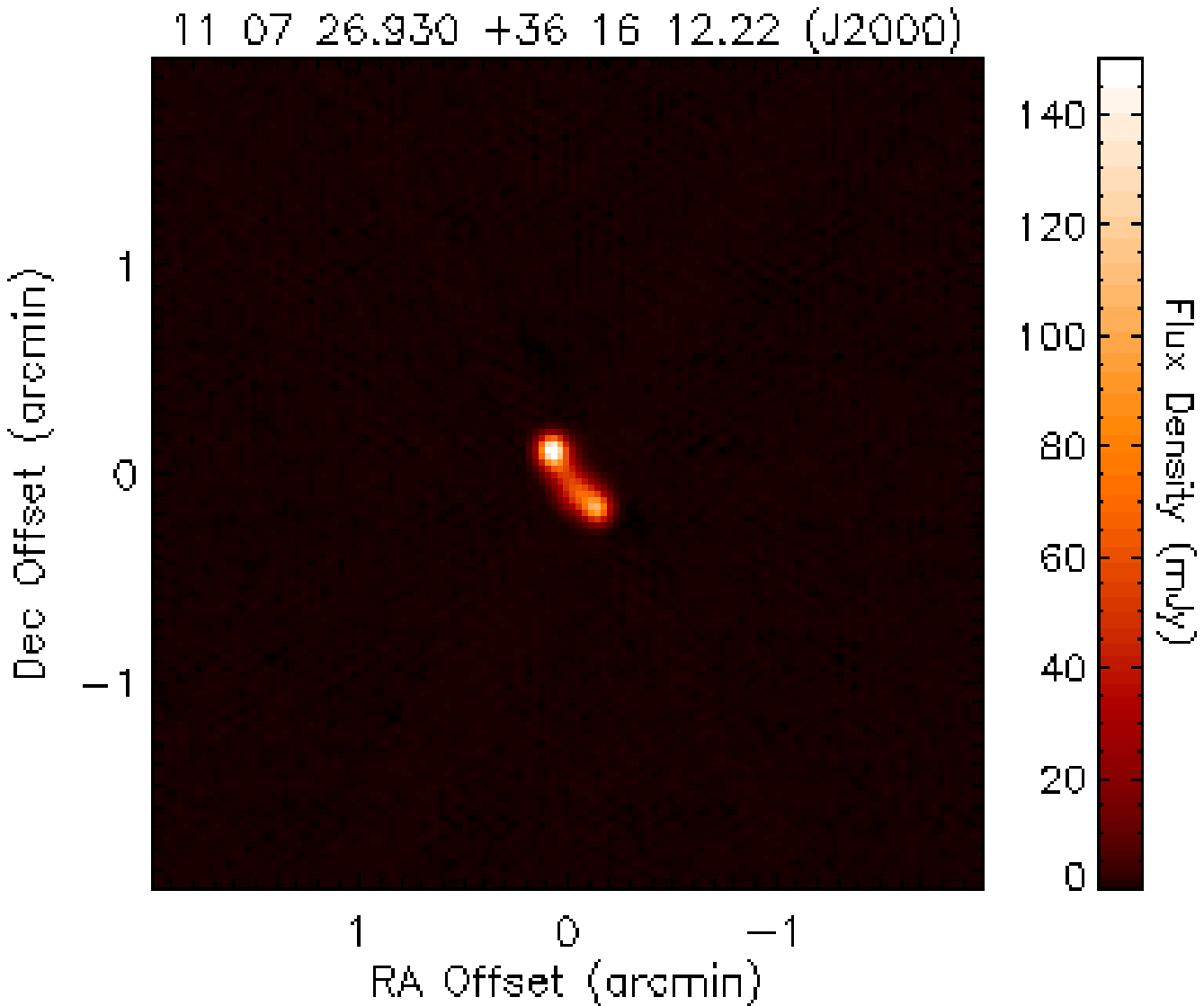}
\includegraphics[width=8.1cm]{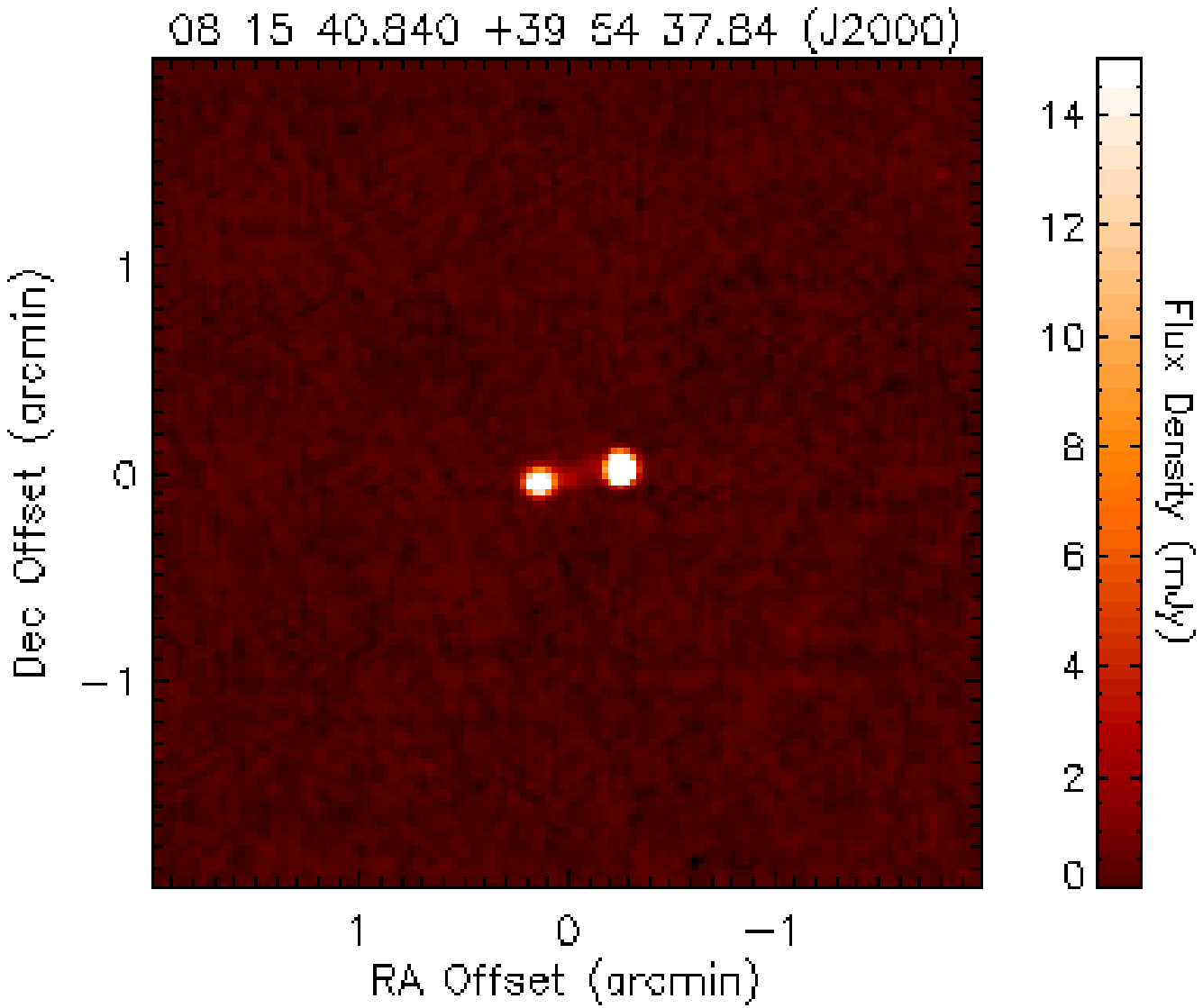}
\includegraphics[width=8.1cm]{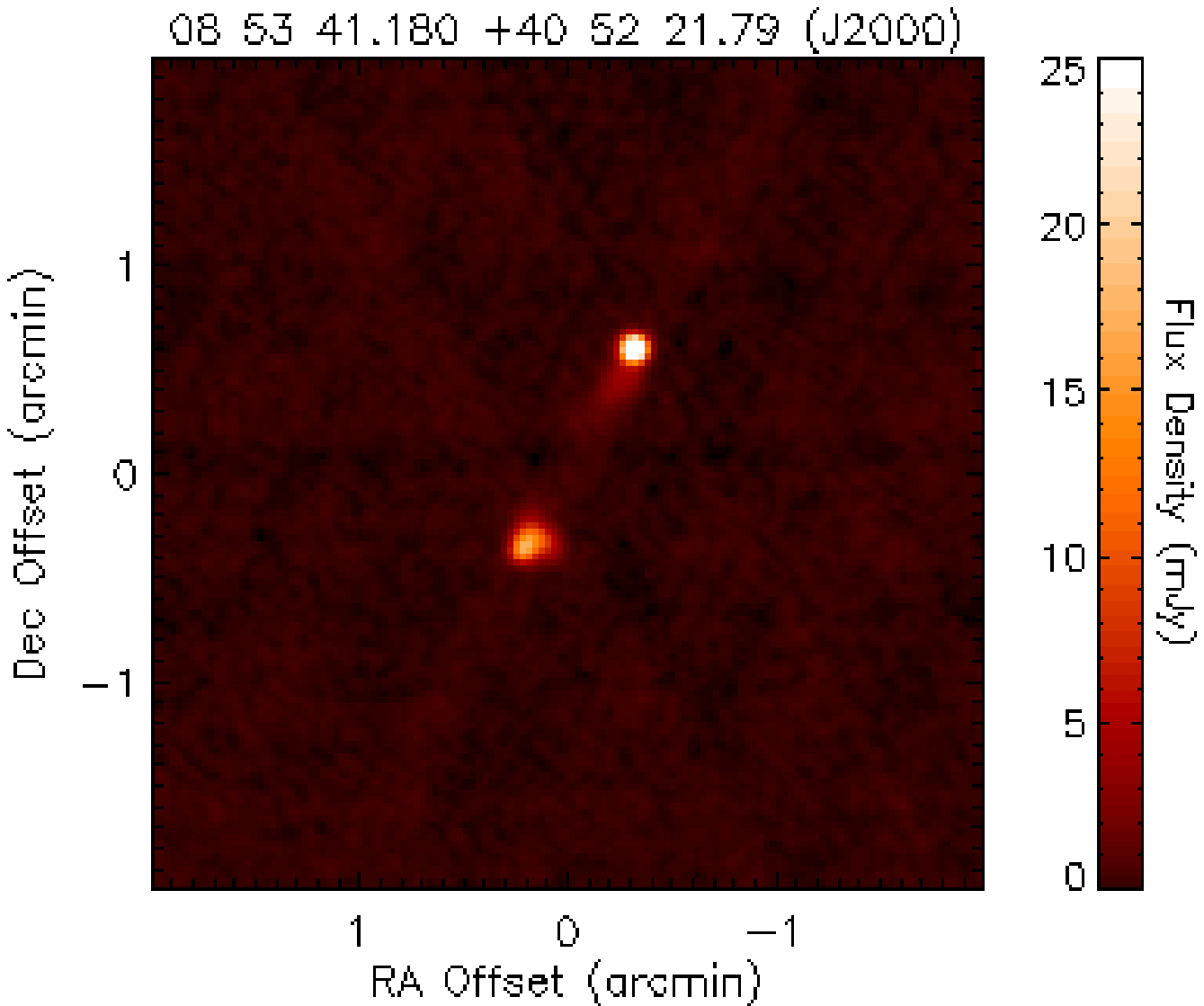}
\includegraphics[width=8.1cm]{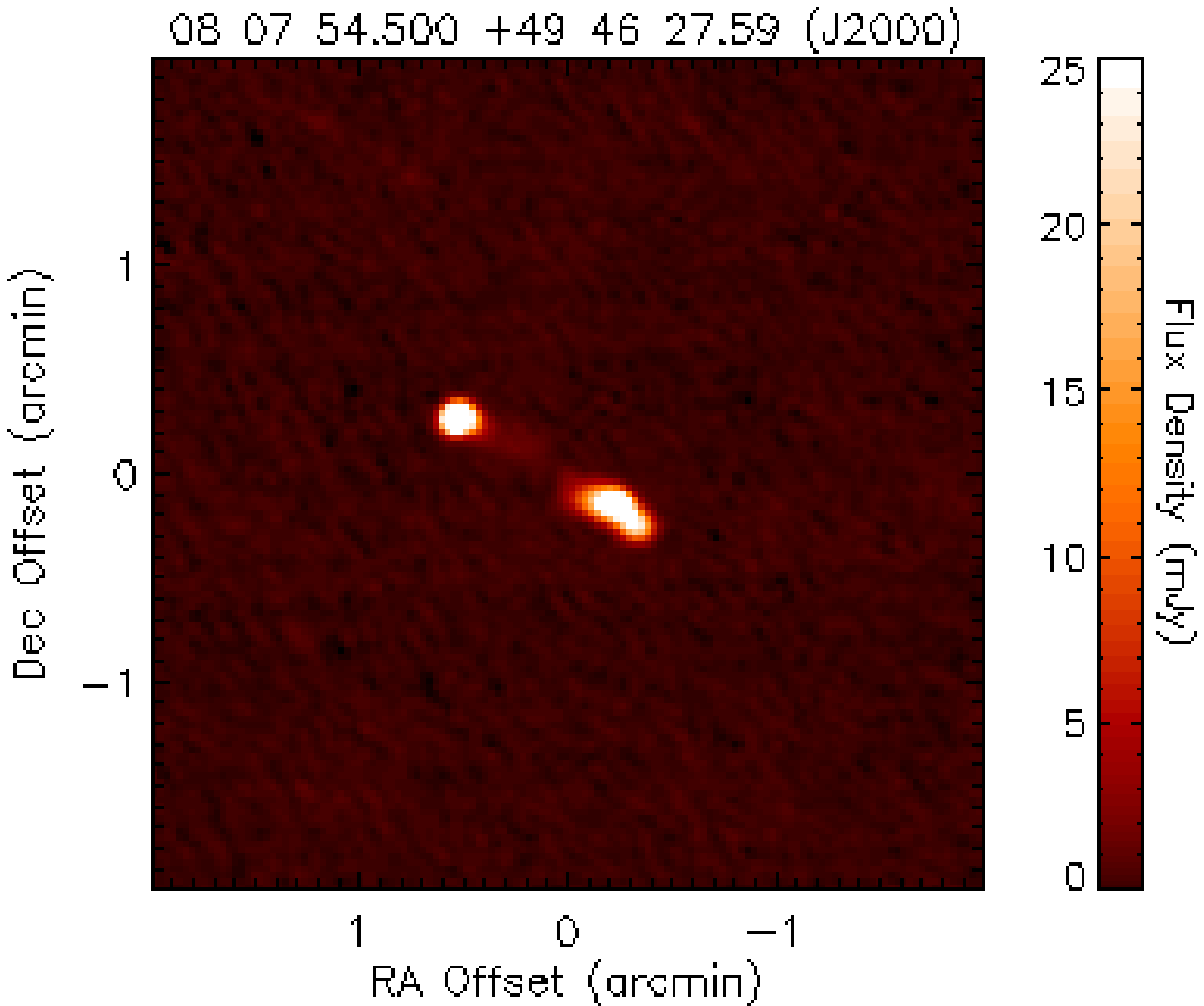}
\caption{FIRST thumbnails showing example of radio morphologies for objects that are selected via our criteria that would not have been selected by matching SDSS and FIRST within 2\arcsec.  From the upper left corner, reading left to right, these objects are J110726.92+361612.2, J081540.84+395437.8, J085341.18+405221.7, and J080754.50+494627.6.\label{fig:firstthumbs}}
\end{center}
\end{figure}

\begin{figure}[htb]
\begin{center}
\includegraphics[width=16.2cm]{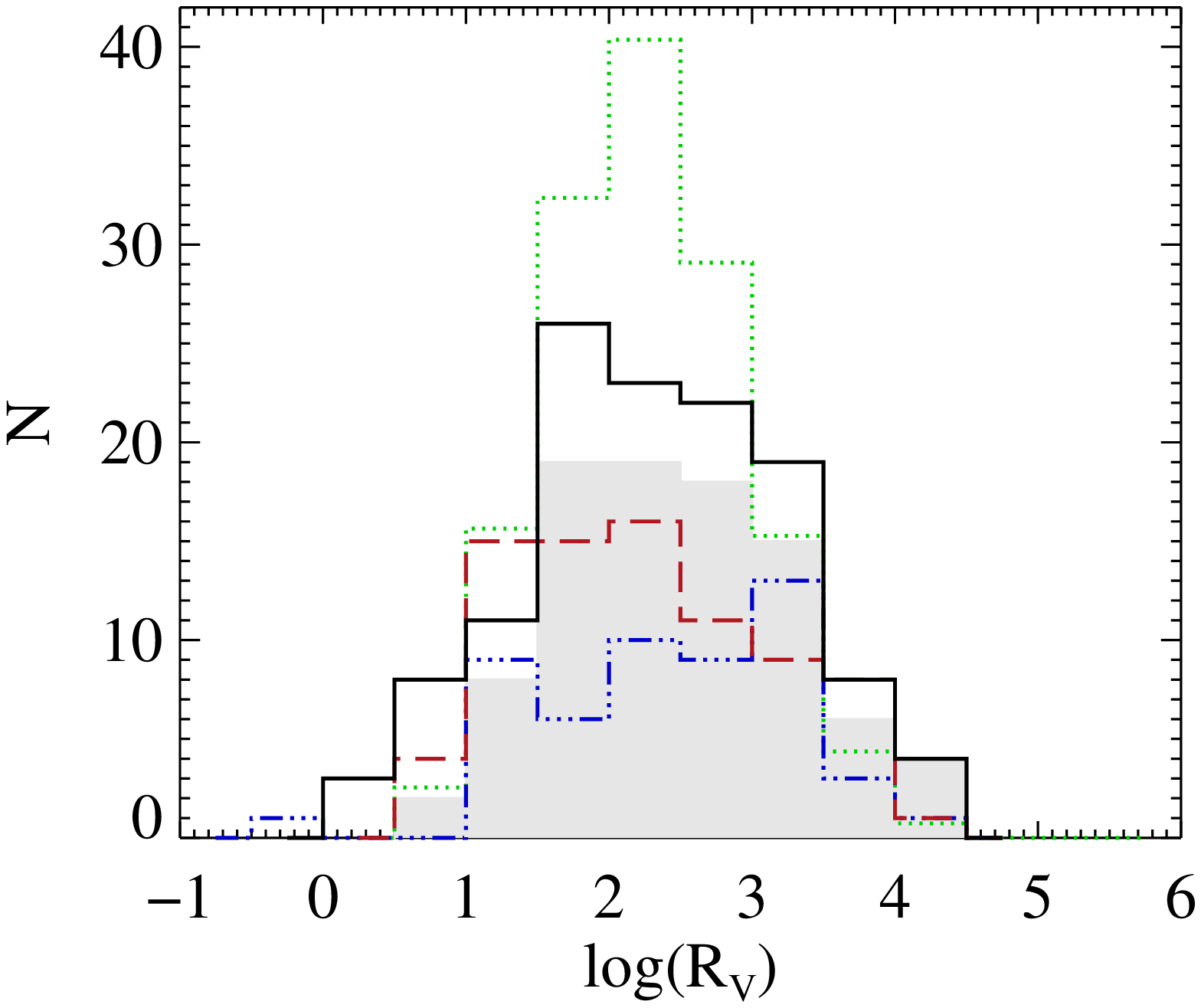}
\caption{A comparison of radio core dominance distributions for the sample compiled in this work (solid black), WB86 (dashed red), SED (dot-dashed blue), and B15 (dotted green). Objects in the new sample that would have been selected by matching our sample with FIRST within 2\arcsec\ are shaded in gray.  The B15 histogram has been divided by a factor of 2.75 for presentation.  Our method notably finds lobe-dominated sources that are very rare or not present in other samples. The difference between the black (this work) and gray (this work, with a FIRST catalog match within 2\arcsec) at $1.5<\textrm{log}(R_{V})<4$ is due to complex matching behaviors including a degeneracy between the FIRST position and source size, see Section~\ref{sec:Rz}.\label{fig:logRhist}}
\end{center}
\end{figure}

We first determined which objects in our sample were selected by our particular set of criteria that would not have been selected by cross-matching SDSS and FIRST within 2\arcsec.  Not surprisingly, there is a noticeable variety in the radio morphologies of the objects that are selected by our criteria.  In particular, we include objects with lobes and no visible core in FIRST that are not present in other samples.  We show some representative examples of the FIRST thumbprints for this population of objects in Figure~\ref{fig:firstthumbs}.  These sources have been previously been preferentially  \citep[but not always, e.g.,][]{devries06,lu07,kimball11} excluded from consideration of orientation effects in quasars.  This translates into a population of low radio core dominance values (log$(R_{V})\lesssim1.5$)  that the straightforward SDSS-FIRST matching methodology does not include (Figure~\ref{fig:logRhist}).  

There is also a population with higher radio core dominance (log$(R_{V})>1.5$) visible in Figure~\ref{fig:logRhist} that would not have been included in the sample by matching SDSS and FIRST within 2\arcsec.  This population of sources hints at more complex behaviors in the sample selection.  An example, SDSS J161742.53+322234.4 is an extended source with log$(R_{V})=2.2$, which falls near the middle of the radio core dominance distribution.  This object is not obtained by matching SDSS and FIRST within 2\arcsec.  The nearest source in the FIRST catalog is clearly associated with the relevant radio structure, but it is 8\farcs2 away.  However, this catalog entry has a deconvolved major axis of 26\farcs46; this illustrates the degeneracy between separation and angular size of the Gaussian fit for each catalog entry.  Other catalogs, for example the DR7 quasar catalog of spectral measurements \citep{shen11} use more complex matching strategies (the first step is to match within 30\arcsec).  The object in question does include a FIRST flux in this catalog, but it is too low by a factor of 4.6 because not all of the FIRST catalog entries were identified.  This more complicated behavior is typical in the sense that there are no clear trends in luminosity, redshift, or angular size that determine which objects would have been missed by matching to FIRST.  It also reflects the known tendency for radio fluxes to be underestimated in $\sim9$\% of quasars by simplistic matching algorithms \citep{lu07}.

\begin{figure}[htb]
\begin{center}
\includegraphics[width=15cm]{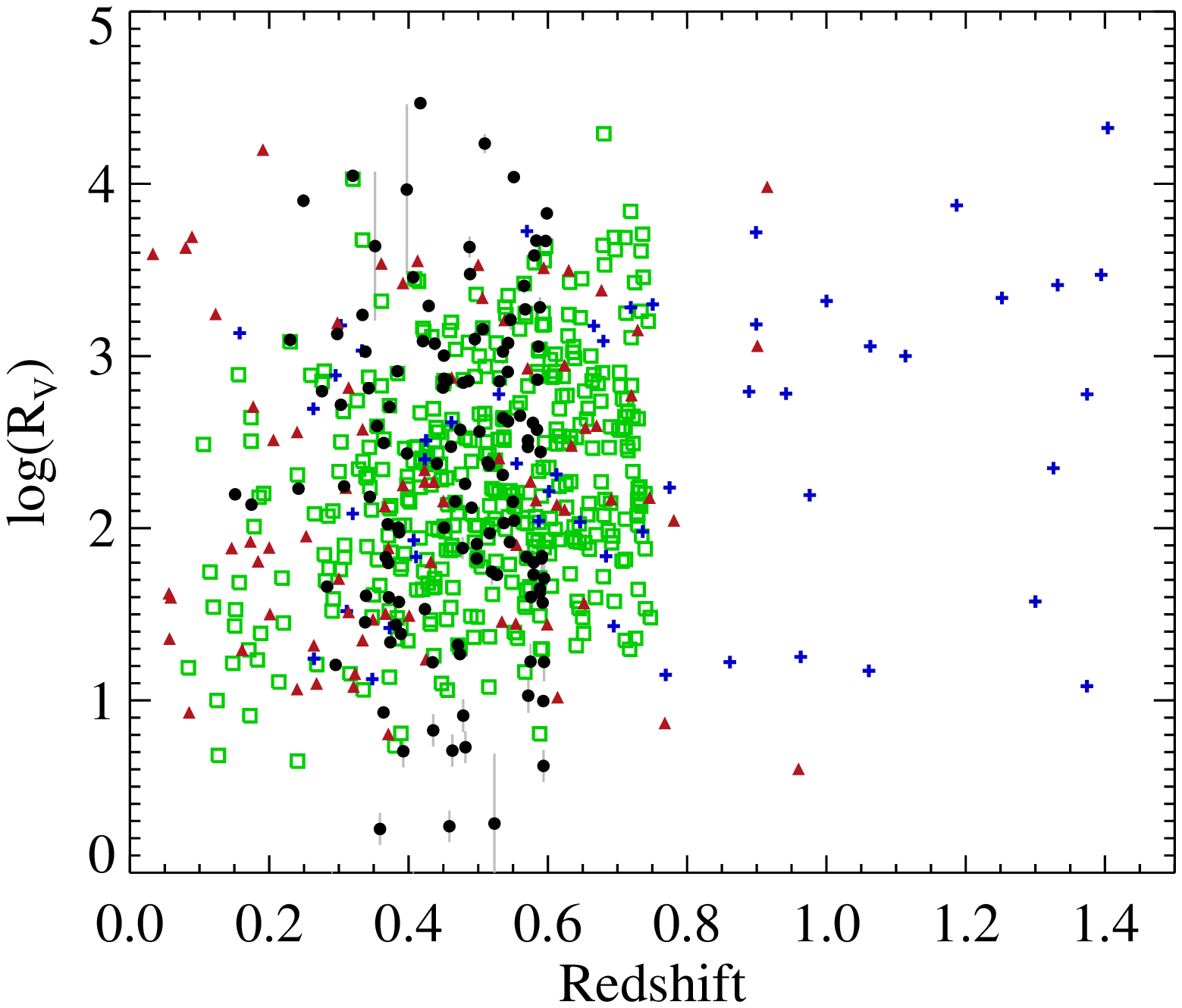}
\caption{Radio core dominance as a function of redshift for the new representative sample (black circles), WB86 sample (red triangles), SED sample (blue crosses), and the B15 (green squares).  Radio fluxes are well known, so uncertainties are almost always smaller than the data points.  The B15 sample has a mild but statistically significant correlation with redshift, and the SED sample is just at the limit of statistical significance. \label{fig:logRz}}
\end{center}
\end{figure}

The redshift dependence from the flux limited samples has crept into the radio core dominance parameter, but the effect is not as strong as the luminosity dependence.  We show this graphically in Figure~\ref{fig:logRz} and numerically in Table~\ref{tab:ks}.  To quantify the degree of correlation, we tabulated the Spearman-Rank correlation coefficients and associated probability of finding the observed distribution of points by chance.  The B15 sample has a mild but statistically significant ($P<0.05$) correlation between radio core dominance and redshift.  The SED sample is just at the limit of statistical significance.

\input{ks_results.tex}

\subsection{Expectation from a uniform distribution of sources with solid angle}
\label{sec:thetadistro}
In a representative sample, the demographics of the radio orientation indicators should match the geometric expectations for a distribution of random orientations.  This means that, for an axisymmetric system like an AGN, jet-on views should be relatively rare compared to equatorial views following the sine of the inclination angle.  We quantitatively compare the observed samples to the radio core dominance distribution expected in this scenario.  

To test this, we adopted the unification by orientation picture where radio-loud quasars and radio galaxies are the same sources viewed at different angles.  We assumed a transition angle between radio-loud quasars and radio galaxies of 60$^{\circ}$, such that quasars will only be observed for $i<60^{\circ}$ \citep{singal93,willott00,marin16,yong20}.  To construct the expected distribution of log$(R_{V})$, we began by uniformly distributing simulated sources in cos$(i)$ and converting these to inclination angles.  The semi-analytic formula from \citet{marin16} provides a relation between inclination and log$(R)$.  It was derived by distributing radio core dominance measurements for the Revised Third Cambridge Catalog of Radio Sources \citep[3CRR,][]{laing83} $z>1$ AGN according to the Copernican Principle (i.e., distributing them evenly in solid angle).  We discuss the implications of this model selection in Section~\ref{sec:discuss}.   We then converted to log$(R_{V})$ assuming log$(R_{V}) = (2.539\pm0.002)+(0.998\pm0.002)\times\textrm{log}(R)$ based on our new representative sample.  Using the relation from \citet{willsbro95} gives qualitatively similar results (but is only calibrated for high values of radio core dominance).  

\begin{figure}[htb]
\begin{center}
\includegraphics[width=15cm]{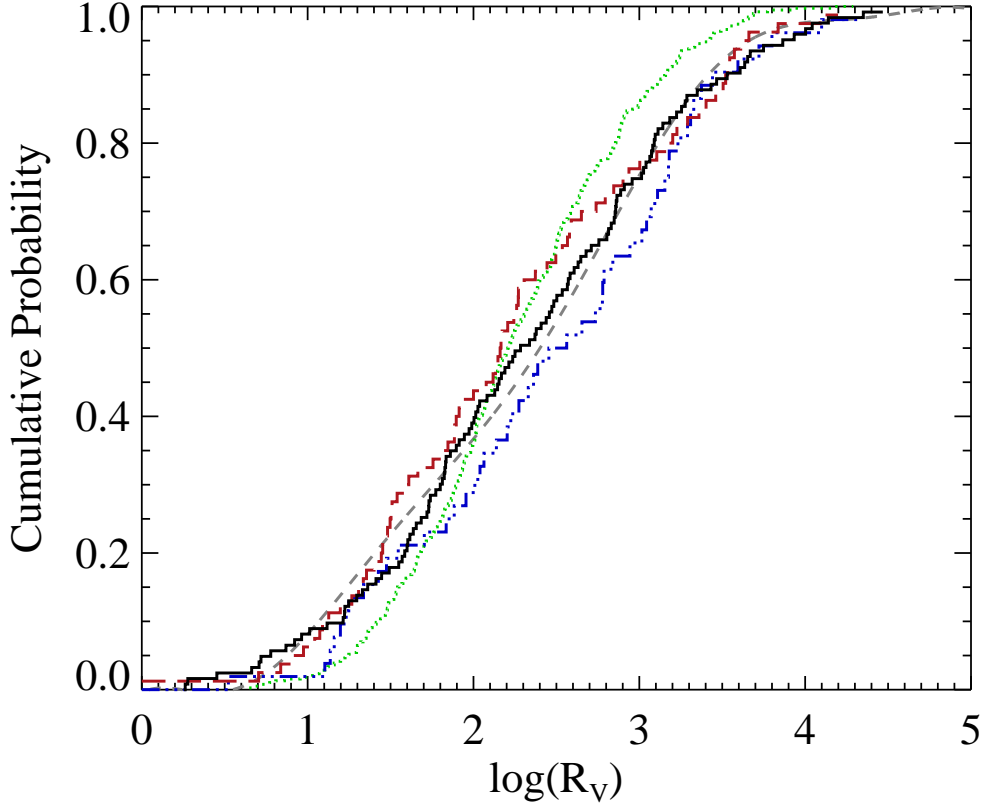}
\caption{The cumulative distribution function for radio core dominance, effectively a ``by-eye'' KS test.  See Table~\ref{tab:ks} for the quantitative results of a one-sample KS test.  The gray dashed line is the expected distribution for sources uniformly distributed in solid angle between $0<i<60^{\circ}$.  The distributions for this work (solid black), WB86 (dashed red), SED (dot-dashed blue), and B15 (dotted green) are overplotted.  The B15 sample is not consistent with being drawn from the expected distribution of log$(R_{V})$ according to this test.  The WB86 sample has the largest divergence from this expected distribution, but is only on the verge of statistical significance due to the smaller sample size. \label{fig:ks}}
\end{center}
\end{figure}

We compare the distributions of log$(R_{V})$ from the observed samples to the expected distribution using a one-sample Kolmogorov-Smirnov (KS) test.  This test quantifies the probability that an empirical distribution is consistent with being drawn from a reference distribution (in this case, the expected log($R_{V}$) distribution).  Specifically, the comparison is made to the cumulative distribution function for the reference function.  Using $10^4$ simulated points we generated the cumulative distribution function for the expected log$(R_{V})$ and reproduced it with a 10th order polynomial at high fidelity.  The D statistics (which characterize the maximum difference in the cumulative distribution for the sample and the reference distribution) and associated probabilities are given in Table~\ref{tab:ks}.  A probability of less than 0.05 indicates that the empirical distribution is not consistent with the expected distribution.  We represent this graphically in Figure~\ref{fig:ks}.  The B15 sample is statistically inconsistent with the expected distribution; it has too many jet-on sources relative to the number of edge-on sources.  The WB86 sample is just at the limit of the stated level of statistical significance, but has the largest divergence from the expected distribution.

There are several important limitations to this comparison of log($R_{V})$ distributions given a conversion between inclination and radio core dominance that warrant further investigation.  The first and most important, is that the expected distribution is sensitive to the selection of transition angle between radio loud quasar and radio galaxy.  This value is not known, but there are observational constraints typically ranging between 45$^{\circ}$ \citep[e.g.,][but see \citealt{dipompeo13d}]{barthel89} and 60$^{\circ}$ \citep{singal93,willott00,marin16,yong20}.  We argue that 60$^{\circ}$ is a reasonable assumption given the evidence from large samples that the value is large, statistical evidence against the smaller estimates \citep{dipompeo13d}, and the fact that smaller values do not produce the low log$(R_{V})$ values observed in all the samples.  

The expected distribution is also sensitive to scatter between inclination and radio core dominance (although less so than the critical angle).  Sources of scatter can include radio variability and, from \citet{willsbro95}, it may be as large as 0.8~dex.  If the scatter is large enough, it obfuscates our ability to discern differences in the radio core dominance distributions, especially with small sample sizes.  The calculation that we perform must therefore be thought of as an illustrative semi-empirical comparison to the geometrical expectation for sources uniformly distributed in solid angle.    

Thirdly, sample size can also play a role, as discussed by \citet{dipompeo13d}.  In the context of constraining the transition angle using samples of radio loud quasars and radio galaxies, they found that the KS test is a conservative way to discriminate between distributions.  That is, for small sample sizes a non-negligible fraction of experiments will not find differences in a sample drawn from a reference distribution.  We characterized the impact of this on our results with a Monte Carlo simulation.  In each of $10^4$ iterations, we drew a sample of N objects that were uniformly distributed in solid angle.  Then we replaced edge-on sources in the sample with random low-inclination values to triple the number of sources with log$(R_{V})>3.5$ (corresponding to log$(R)>1$ or $i<18^{\circ}$) while keeping the total sample size the same.  Thus, each distribution was modified from the geometric expectation and we stored the results of a one-sample KS test.  The SED sample and the sample derived in this work were simulated by adopting $N\sim50$ and $N\sim100$.  We find that the one-sample KS test detects a statistically significant difference with $P<0.05$ less than 87\% of the time for a small SED-like sample, whereas the difference is detected $>99.9$\% of the time for the larger sample consistent with this work.  Of course, the outcome of this exercise is sensitive to the magnitude of the difference and other details of the analysis.  Therefore, another way of phrasing the lesson learned from this exercise would be to say that in a sample as small as the SED sample, the KS test is can only recover large differences from geometric expectation.

\section{The orientation dependence of quasar broad emission lines}
\label{sec:analysis}
Having established the merits of our sample selection methodology and the credentials of the resulting representative sample, we employed it to explore orientation effects in quasars.  

\begin{figure}[htb]
\begin{center}
\hspace{-1.45cm}
\includegraphics[width=10.95cm]{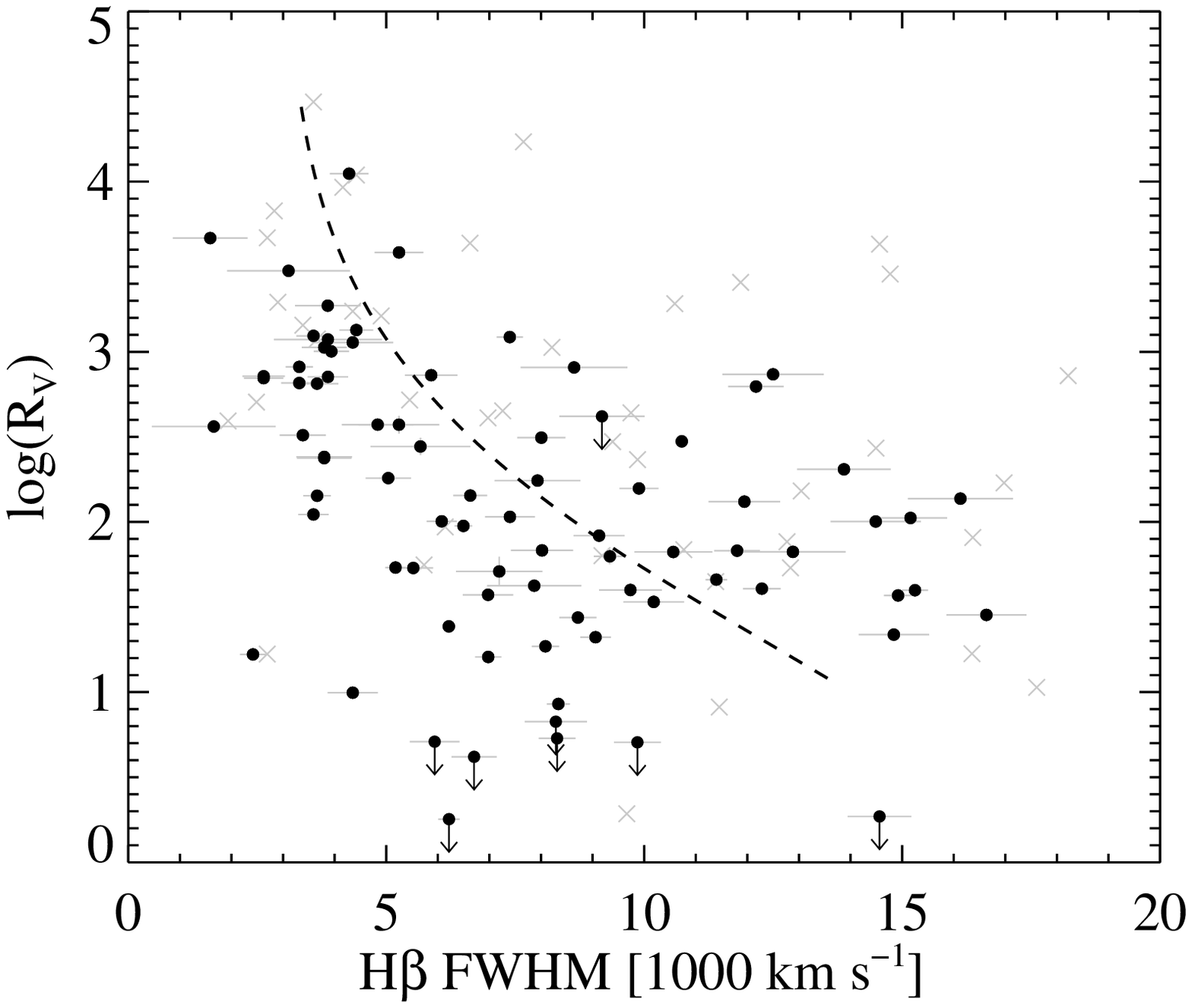}
\hspace{-2.75cm}\includegraphics[width=10.95cm]{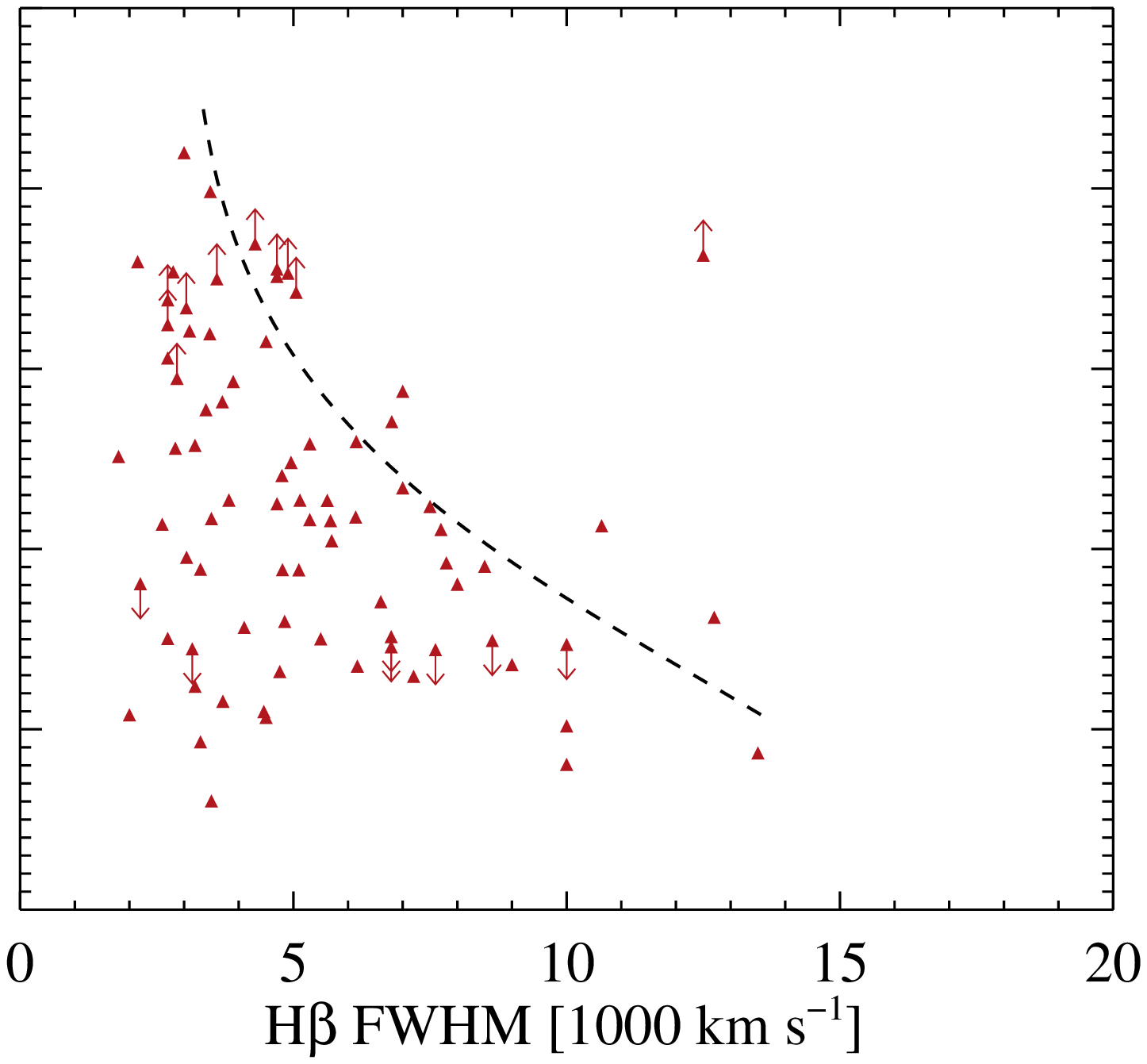}
\caption{The orientation dependence of the broad H$\beta$ line for this sample (left) and the WB86 sample (right). Arrows indicate limits on radio core dominance, in the new sample this occurs when no core component was detected in FIRST.  Uncertainties in the log$(R_{V})$ direction are generally smaller than the data points because the radio fluxes are well known.  Gray crosses show objects in our sample with low-quality radio or optical measurements (see Section~\ref{sec:analysis}).  The beaming model (shown by the dotted line) is translated to log$(R_{V})$ using the relationship in \citet{willsbro95}.  The historical orientation dependence for \Hb\ is reproduced, but with significantly more scatter, and a population of objects above the dashed model curve that was not present in the WB86 sample.\label{fig:wb86}}
\end{center}
\end{figure}

In Figure~\ref{fig:wb86} we show the radio core dominance orientation indicator, log$(R_{V})$, against the FWHM of the broad \Hb\ emission line for the new sample and WB86.  Although we show all 123 objects with both measurements, we highlight the objects with high-quality measurements indicating that their location in the diagram is reliable.   These are the 72 spatially resolved sources with angular sizes in FIRST of $>\!5$\arcsec\ where the peak of \Hb\ is at least $3\sigma$ above the noise and we can make reliable FWHM measurements.   

The distribution of points in this space can be compared to the same data from \citet{wills86}, which we reproduce assuming the \citet{willsbro95} conversion to log$(R_{V})$ in the right panel of the figure.  The measured values of log$(R_{V})$ are not available, but can be viewed for the same sample in \citet{willsbro95}.  We find a similar distribution as those works, but the envelope is not as sharply defined.  The difference has been noted before \citep{brotherton15} and is particularly striking; in the WB86 sample only 10/79 (12\%) objects fall above (or to the right of) the beaming model which characterizes the envelope edge well, whereas we find that 29/89 (32\%) objects in our sample fall above this line when it is translated to log$(R_{V})$ assuming the \citet{willsbro95} conversion from log$(R)$.  We point out that there are two possible ways of framing this increased scatter, because points can be thought of as falling above the model line or to the right of it depending on the origin of the offset.  We will revisit the details of the model in Section~\ref{sec:wb86r}, but first we explore the data and measurements in a model-free way.

In an effort to find the zeroth order difference between our dataset and that of \citet{wills86}, we looked first to the sample selection.  In particular, we tried to identify mechanisms that might place objects above the sharp \citet{wills86} envelope (typically core dominated objects with broad emission lines), as these seem to represent the biggest difference between the samples.  The most obvious consideration, that these objects were added by our new selection criteria, is ruled out because our method has added primarily very lobe-dominated sources to the sample (see Figure~\ref{fig:logRhist}).  These are not exclusively (or even primarily) sources in our sample with no FIRST match within 2\arcsec, either.

\begin{figure}[htb]
\begin{center}
\includegraphics[width=16.2cm]{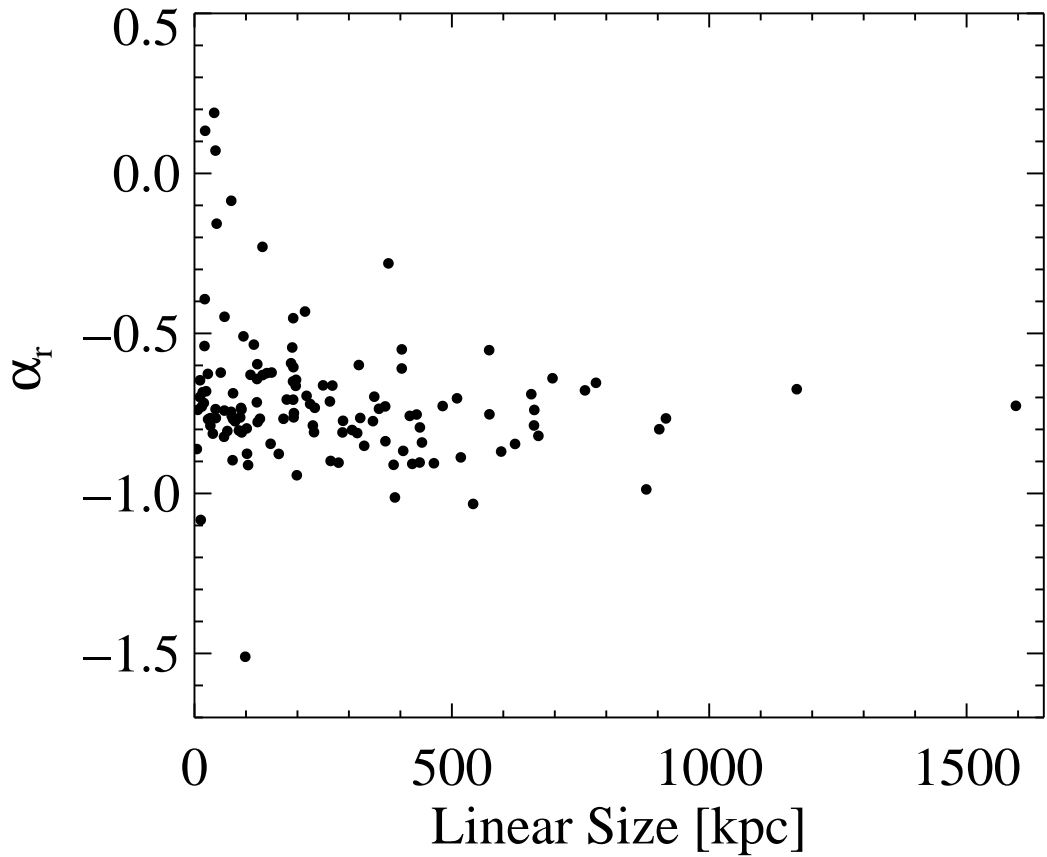}
\caption{Radio spectral index versus the largest linear size for all 126 sources.  CSS sources with compact linear sizes ($d<20$~kpc) and steep radio spectra ($\alpha_{r}<-0.5$) may have over-estimated radio core dominance measurements, thus smearing the sharp envelope in FWHM--log$(R_{V})$ space.  However, these objects were excluded from the analysis by the resolution cut at 5\farcs0 to remove things unresolved by FIRST.  See Section~\ref{sec:analysis} for details. \label{fig:css}}
\end{center}
\end{figure}

Another possibility raised by \citet{brotherton15} is that a population of core-dominated, broad-lined objects was added by the inclusion of compact steep-spectrum (CSS) sources.  These objects have compact radio morphologies and are therefore assigned large values of radio core dominance, but their steep spectra suggest that if their radio structures could be resolved, they would yield smaller values of $\textrm{log}\left(R_{V}\right)$.  Specifically, CSS sources are typically characterized as having compact linear sizes ($d<20$~kpc) and steep ($\alpha_r<-0.5$) spectra \citep[e.g.,][]{odea98}.  In fact, at the redshifts of our sample the resolution cut of 5\farcs0 that we applied when identifying high-fidelity measurements removed sources with linear sizes $<30$~kpc, so CSS were already effectively excluded.  However, as an interesting aside, Figure~\ref{fig:css} shows a continuous distribution of radio spectral index with largest linear size (which is calculated from the largest angular size determined from FIRST), implying that CSS sources are not a distinct population.  Thus, if they are indeed young sources the distribution of ages must also be continuous.

Although the additional population of objects is not caused by CSS sources, there are a number of sample members with large values of radio core dominance but steep radio spectra.  We investigated this further by visually inspecting the FIRST maps and any archival data for all the sources with log$(R_{V})>2.9$ (roughly corresponding to log$(R)>0.5$) and $\alpha_{r}<-0.5$.  There are 27 such objects.  By considering all the available radio data, including maps at other frequencies and higher resolution as well as fluxes at a variety of wavelengths, we saw several common behaviors.  A substantial fraction of objects appear to truly have steep radio spectra and point-like morphologies in FIRST.  Another subset of objects have complex radio spectra, which may not be well characterized by a single slope.  Finally, there are some sources where the radio core dominance measurement appears to be robust and the slope we have measured does not match up with a radio spectrum that includes data at more wavelengths.  The takeaway from this is that there is likely some contamination where the radio orientation parameters break down, not surprising given their statistical nature, but this cannot explain the addition of all the points above the sharp \citet{wills86} envelope.  

Finally, we considered whether physical properties might cause horizontal scatter in the FWHM--log$(R_{V})$ diagram.  That is, we consider whether the new population of objects falls to the right of the sharp envelope, rather than above it.  The additional scatter cannot represent the jet-on view of a population of objects having very broad emission lines (presumably as a result of intrinsically higher black hole masses), for lack of a corresponding population of objects viewed from an edge-on orientation.  Statistically, the edge-on viewing angle is much more probable than the jet-on one, so this explanation is disfavored because we do not see edge-on sources with significantly broader lines than \citet{wills86}.  Although the population of sources does have fairly high single-epoch black hole mass estimates, commensurate with their broad line widths, they span the full range observed in the sample and the radio core dominance distribution for this population appears biased towards high values compared to geometric expectations.  

\begin{figure}[htb]
\begin{center}
\includegraphics[width=16.2cm]{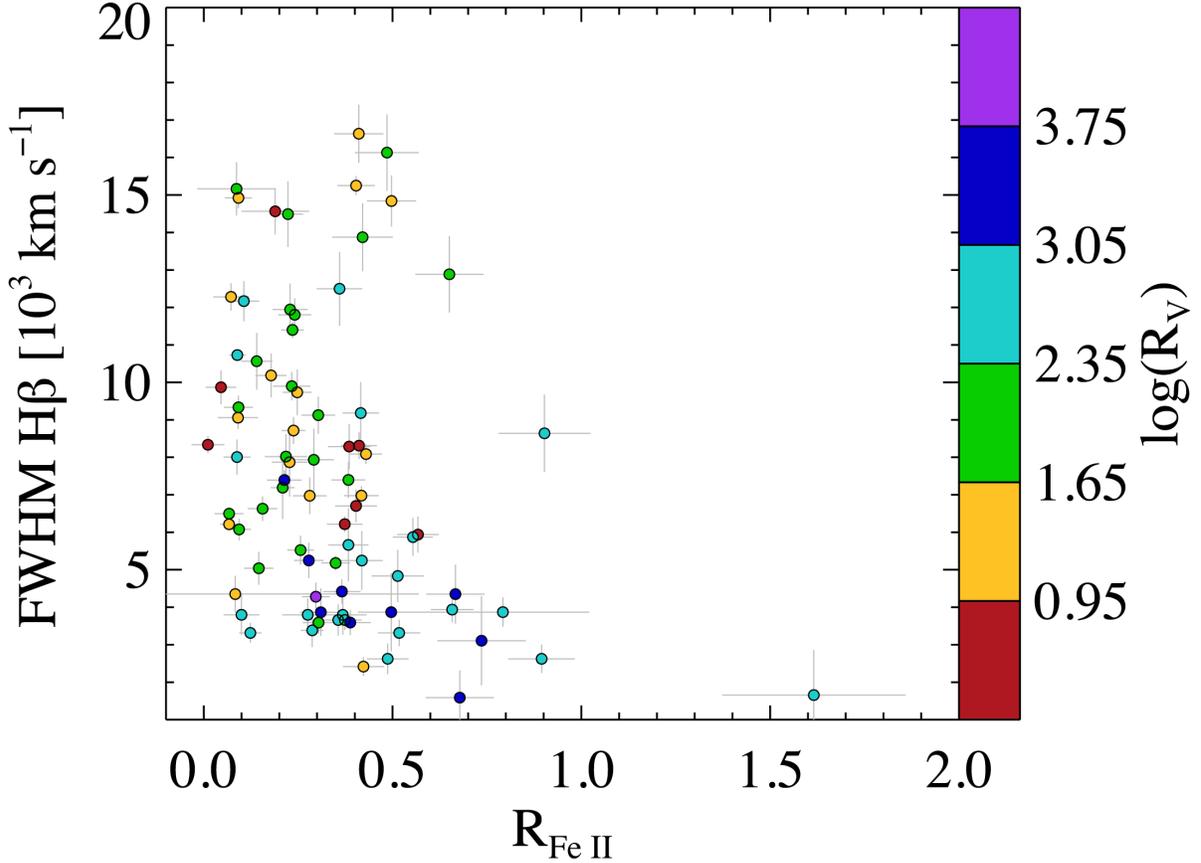}
\caption{The FWHM of \Hb\ versus $R_{\textrm{\FeII}}=\textrm{EW}(\textrm{\FeII}/\textrm{\Hb})$.  Here, $R_{\textrm{\FeII}}$ is taken to be a proxy for the Eddington fraction and we investigate the possibility that, for a given value of $R_{\textrm{\FeII}}$, the spread in the FWHM of \Hb\ is caused by orientation.  In this sample, there is no statistically significant evidence for such an orientation dependence.    \label{fig:shen14}}
\end{center}
\end{figure}

Alternatively, Eddington ratio may introduce scatter into the FWHM--log$(R_{V})$ diagram.  The idea that the diversity of quasar spectral properties can be unified with orientation and accretion has been suggested multiple times in the literature \citep[e.g.,][]{marziani01,shen14}.  The claim is that, at a given value of $R_{\textrm{\FeII}}$, a proxy for the Eddington ratio, spread in the FWHM of \Hb\ is due to orientation.  In Figure~\ref{fig:shen14}, we show the FWHM of \Hb\ versus $R_{\textrm{\FeII}}$ color coded in bins of log$(R_{V})$.  The core dominated objects all have noticeably small values of the FWHM, as is expected from the distribution of objects in FWHM versus log$(R_{V})$ space.  But beyond this there is considerable scatter in the diagram and it is not clear that orientation effects are responsible or the range in FWHM at a given value of $R_{\textrm{\FeII}}$.  Or, conversely, $R_{\textrm{\FeII}}$ does not clearly determine an object's location in FWHM--log$(R_{V})$ space.  More sophisticated statistical analyses including correlations and clustering in measurement space do not add any new information to this conclusion.  This brings an interesting point into the spotlight: quasar spectra are very diverse and can be unified by orientation and accretion only in the broadest sense.  The clear trends of \OIII\ EW in FWHM versus $R_{\textrm{\FeII}}$ space illustrated by \citet{shen14} are only visible when the data are smoothed.  This means that these underlying trends are hidden by substantial spectral differences from quasar to quasar that have not been accounted for.  As a corollary to this, it is possible that with a significantly larger sample it would be possible to tease out the expected orientation dependence of FWHM at fixed $R_{\textrm{\FeII}}$.  We note here that the picture for radio-quiet (RQ) quasars may be different.  RL objects typically lie on one end of the EV1 distribution \citep[e.g.,][]{bg92,richards11}, and our sample covers less than an order of magnitude in $R_{\textrm{\FeII}}$.  That said, our sources do cover the whole range shown in figure~1 of \citet{shen14} (which includes RQ objects).  

\section{An updated model}
\label{sec:wb86r}

\citet{wills86} demonstrated the difference in H$\beta$ line width for high and low radio core dominance source can be attributed to viewing the same system at different inclinations, instead of intrinsic differences between those populations.  Their model for the BLR, which we show in Figure~\ref{fig:wb86}, is consistent with orbiting, infalling, or outflowing emission-line gas that is predominantly in the plane of a disk perpendicular to the radio axis.  The observed velocity is a combination of a random isotropic component, $v_r$, and a component that is in the plane of the disk, $v_p$, given by:

\begin{equation}
\label{eqn:blr}
(v_r^2 + v_p^2\, \textrm{sin}^2\theta )^{1/2}.
\end{equation}

Their original envelope in the data which we reproduce in Figure~\ref{fig:wb86} is described with $v_r = 4,000$~km~s~$^{-1}$ and $v_p = 13,000$~km~s~$^{-1}$.  The angle, $\theta$, between the radio axis and the observer's line of sight is related to radio core dominance by special relativity in the simple case where the compact, relativistically beamed radio core results the unresolved bases of two opposite radio jets with a single bulk velocity.  Originally laid out by \citet{scheuer79}, \citet{wills86} followed equation 3 of \citet{orr82} which includes a contribution from the receding side of the compact core.  In this model, radio core dominance is a function of inclination angle, $R_T = R(90^{\circ})$, and the Lorentz Factor, $\gamma$.  The parameters $R_T$ and $\gamma$ effectively determine the minimum and range that radio core dominance can have.  \citet{orr82} estimated $\gamma=5$ and $R_T=0.024$ from the observed distribution of radio core dominance in a sample of radio-loud quasars.    

We explored whether the model of \citet{wills86} is able to describe our dataset if the values of the model parameters are updated.  Notably, the \citet{orr82} model for the radio quasar does not reproduce measured values of radio core dominance in more recent samples (including ours).  The distribution of radio core dominance is too narrow, even for values of $\gamma$ and $R_T$ that produce the correct range and minimum.  This was previously noted by \citet{marin16}.  In comparison, their semi-empirical model which we adopted previously in this work is geometrically motivated (i.e. to distribute sources evenly in solid angle) and produces more core fluxes at intermediate angles.  Their broader distribution can be attributed to divergence from a single-zone jet with zero opening angle, potentially by variety in bulk speed or optical depth \citep{blandford79}.  

\begin{figure}[htb]
\begin{center}
\includegraphics[width=16.2cm]{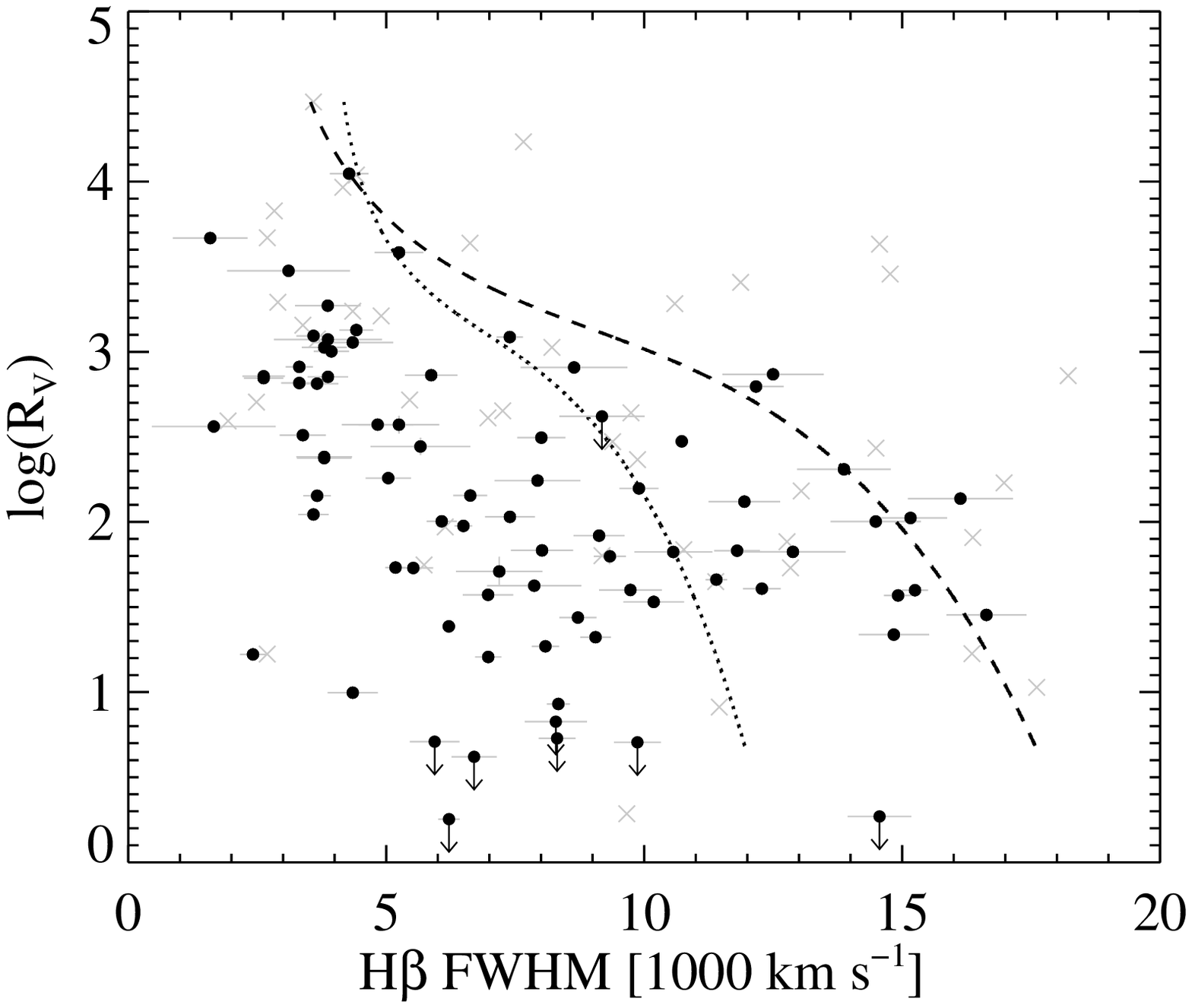}
\caption{The orientation dependence of the broad H$\beta$ line for this sample with updated beaming models.  The dotted line shows the \citet{wills86} beaming model with $v_{r}=4,000$~km~s$^{-1}$ and $v_{p}=13,000$~km~s$^{-1}$, but with the \citet{marin16} relationship between inclination angle and radio core dominance.  The dashed line is the same, but has $v_{r}=3,000$~km~s$^{-1}$ and $v_{p}=20,000$~km~s$^{-1}$.  Arrows indicate limits on radio core dominance.  Uncertainties in the log$(R_{V})$ direction are generally smaller than the data points because the radio fluxes are well known.  Gray crosses show objects in our sample with low-quality radio or optical measurements (see Section~\ref{sec:analysis}).    \label{fig:wb86r}}
\end{center}
\end{figure}

In Figure~\ref{fig:wb86r}, we show updated models for quasar beaming.  These are in the spirit of the \citet{wills86} model, but include the \citet{marin16} prescription for observing the inclination angle.  We show one version of the model where this is the only change from \citet{wills86}, and another where we adopt $v_r = 3,000$~km~s~$^{-1}$ and $v_p = 20,000$~km~s~$^{-1}$.  The latter describes the envelope observed in our dataset.  

Ultimately, the major difference between our dataset and the model as it was originally calculated in \citet{wills86} is with the diversity of radio jet properties.  However, the parameters describing the velocity field subsequently determine what part of the FWHM--log$(R_{V})$ space is filled and the sample selection determines the location and sharpness of any envelope in that space.

\section{Discussion}
\label{sec:discuss}
In this work, we present a representative sample of 126 $0.1<z<0.6$ AGN with log$(L_{325})<33.0$ erg~s${-1}$~Hz$^{-1}$.  We demonstrate that biases resulting primarily from flux limited, small matching radius, high radio frequency selection are mitigated compared to several samples from the literature \citep{wills86,shang11,brotherton15}.  Additionally, we highlight a population of objects with relatively jet-on viewing angles but broad optical emission lines and explore the origin of these sources.   Our main result is that this sample can be described by the physical picture for quasar beaming presented by \citet{wills86} only if the diversity in the velocity field and radio jets is larger than previously required.


Our approach to building a representative sample of quasars for studying orientation uses selection criteria that are already prevalent in the literature.  Low-frequency selection to target radio lobes was common in early orientation samples, for example in the 3CRR catalog.  Additionally, there are many sophisticated examples of catalog cross-matching procedures qualitatively identical to the one that we adopt, although applied to FIRST where radio lobes are more difficult to detect because they are extended and have steep spectra \citep{devries06, lu07, kimball11}.  Thus, what is new in our selection is the combination of low radio frequency selection, sophisticated cross matching, and a low-frequency radio luminosity cut.

One limitation associated with this work is that the radio orientation indicators were measured from survey quality data.  Timing effects may be at work, either in the form of variability between observations at different wavelengths which were not taken simultaneously or in the sense that variations in the jet may not yet have propagated to large scales.  This may serve to degrade the correlation between radio core dominance and orientation angle in individual objects.  \citet{jackson13} detail the effects that FIRST resolution and sensitivity limits can have on orientation studies in AGN.  Among their findings are that a substantial number of objects with lobes and no core will be missed due to the sensitivity limit, core dominated objects will be under represented due to limited resolution when visual identification of a lobe is required to measure radio core dominance \citep[e.g.,][]{kimball11}, and that radio core dominance values can be boosted when lobe emission is accidentally attributed to the core.  One particular result of these effects is that correlations with orientation that were found in early works \citep[e.g.,][]{jackson89} will appear weaker in large survey samples \citep[e.g.,][]{kimball11} because it is the core-dominated sources that drive the correlation to stand out.  Some of these issues are mitigated in our sample.  Notably, our particular sample selection allows us to identify and include the sources with lobes and no core and we have made an effort to use the largest angular size in FIRST to identify unresolved sources where radio core dominance measurements are potentially unreliable.  The latter choice may remove some core dominated objects, however, in a sample unbiased by orientation these should be rare.  Nevertheless, new high-resolution radio observations of the most compact objects would be the ideal way to refine the orientation indicators for this sample and learn what, if any, effect this has in producing the population of objects with high log$(R_{V})$ and FWHM (Maithil et al. in prep.).

Although radio core dominance is clearly an orientation indicator \citep{ghisellini93}, the exact relationship between the the two is not well established over all angles.  The \citet{marin16} prescription that relates inclination angle to radio core dominance has important limitations.  Namely, it is based on the observed radio core dominance values in the 3CRR $z>1$ AGN, which have been associated with inclinations by distributing them evenly in solid angle.  Given the importance of sample selection outlined in this work, it is worth noting that the 3CRR catalog is flux limited to 10.9~Jy at 178~MHz.  The low-frequency selection is a plus, but the limitations of flux-limited samples apply and such a high flux limit relegates members to the very highest luminosities at each redshift.  Alternative empirical relations \citep[e.g.,][]{willsbro95, dipompeo12a} are limited by the small number of sources with inclination measurements and physical models \citep{orr82} are limited by simplistic (e.g., single zone jet) assumptions.  \citet{marin16} argue that more sophisticated physical models, like popular spine-sheath models where the high-velocity jet spine has a lower-velocity sheath \citep[e.g.,][]{sol89} would be more consistent with their semi-empirical model.  Despite this uncertainty, the exercises we have conducted still yield useful lessons.  Namely, the radio core dominance distribution in representative samples (and even large survey samples) is broader and includes more intermediate objects than in samples from the 1980s.  The physical picture of the BLR from \citet{wills86} may still describe the observations, but more diverse jet and BLR properties need to be reflected in the model parameters.  It is not clear what if any selection or other observational property (e.g., $R_{\textrm{\FeII}}$) drives this.


\section{Summary}
\label{sec:summary}
This work highlights the properties of a sample of AGN with radio observations and optical spectra that is selected independent of orientation.  In particular, we used a sophisticated procedure to match members of the SDSS DR7 quasar catalog with $0.1<z<0.6$ to WENSS, a 325~MHz survey where anisotropic beaming effects are minimized, and applied a total radio luminosity cut of log$(L_{325})<33.0$ erg~s${-1}$~Hz$^{-1}$.  The result is a sample of 126 AGN with optical spectra and radio observations which is generally representative of the larger population of quasars.

Our conclusions are as follows:

\begin{itemize}
\item In the context of previously selected samples in the literature  \citep{wills86,shang11,brotherton15}, this sample is less biased.  Specifically, at every redshift it includes objects down to the limiting luminosity, rather than the most luminous objects at every redshift as in previous samples.  Our selection criteria add sources with lobe-dominated morphologies over a selection that matches the SDSS to the larger area FIRST survey at 1.4~GHz.  In practice, some of these objects have unusual radio morphologies at 1.4~GHz, for example radio lobes but no core.  Additionally, the distribution of radio core dominance in this sample is consistent with the expectation for sources distributed uniformly in solid angle between 0 and 60$^{\circ}$ \citep{marin16}, unlike samples from \citet{wills86} and \citet{brotherton15}.

\item Emphasizing sample members with the highest quality measurements, we looked for relationships between radio core dominance and the velocity widths of the broad \Hb\ emission line.  We recovered the dependence observed by many before \citep[e.g.,][]{wills86}, where likely jet-on are limited to narrow H$\beta$ widths and those that are likely edge-on show a larger range of H$\beta$ widths, however with significantly more scatter.  Notably, the sharp envelope that was previously observed in FWHM--log$(R_{V})$ space is absent from our sample due to scatter along one or both axes.  We looked for contamination by CSS sources, variation in black hole mass or EV1, or differences among the sample selection but were not able to identify a single mechanism that would introduce this difference.

\item The \citet{wills86} physical framework can describe the representative sample if the parameters of the model are diversified.  By adopting a semi-empirical prescription for beaming \citep{marin16} rather than a physical model for a single-zone jet \citep{orr82}, and increasing the maximum values in the BLR velocity field, we find a new envelope that is consistent with the data.  This primarily reflects the need for higher velocities and a departure from a simple, single-component jet model to one that allows for diversity in jet properties like variation in bulk speed, opening angle, or optical depth effects.  The takeaway is that the physical picture of quasar BLRs laid out by \citet{wills86} holds, but biases in sample selection have the ability to create the expectation that only a small range in physical parameters are necessary to explain the data.
\end{itemize}

In the future, this sample will be useful for ongoing investigations of quasar orientation indicators and testing more complex models for quasar geometry.  For example, \citet{brotherton15} demonstrates that the difference in black hole mass derived from \Hb\ FWHM and stellar velocity dispersion may provide a radio-quiet orientation indicator.  The added dynamic range in radio core dominance by going to the most edge-one orientations will be useful in the development of this concept.

\acknowledgements{This work was made possible by NGS grant \#$9471-14$ and a contribution from the NASA Pennsylvania Space Grant and was performed in part at the Aspen Center for Physics, which is supported by National Science Foundation grant PHY-1607611.  JCR would like to thank Bev Wills, Kayhan G{\"u}ltekin, Rajib Ganguly, Mike Brotherton, Mike Eracleous, and Mike DiPompeo, for helpful discussions during the preparation of this work.  

This research has made use of the NASA/IPAC Extragalactic Database (NED), which is operated by the Jet Propulsion Laboratory, California Institute of Technology, under contract with the National Aeronautics and Space Administration.
}


\bibliographystyle{aasjournal}
\bibliography{all.052615}

\label{lastpage}
\end{document}

%% file: apj_commands.tex
\newcounter{species}

\newcommand{\OIIIdblt}{[O\,{\sc iii}]~$\lambda\lambda$4959, 5007}

\newcommand{\OIIIw}{[O$\,\textsc{iii}]$~$\lambda$5007}

\newcommand{\Hb}{{H}$\beta$}

\newcommand{\FeII}{Fe\,{\sc ii}}

\newcommand{\OIII}{[O\,{\sc iii}]}

\def\lsim{\lower0.3em\hbox{$\,\buildrel <\over\sim\,$}}
\def\gsim{\lower0.3em\hbox{$\,\buildrel >\over\sim\,$}}

\def\sun{\hbox{$\odot$}}
\def\earth{\hbox{$\oplus$}}
\def\lesssim{\mathrel{\hbox{\rlap{\hbox{%
 \lower4pt\hbox{$\sim$}}}\hbox{$<$}}}}
\def\gtrsim{\mathrel{\hbox{\rlap{\hbox{%
 \lower4pt\hbox{$\sim$}}}\hbox{$>$}}}}

\def\arcsec{\hbox{$^{\prime\prime}$}}

\def\farcs{\hbox{$.\!\!^{\prime\prime}$}}

%% file: all_prop_short.tex
\begin{deluxetable}{lccccccccccc}
\setlength{\tabcolsep}{4pt}
\tabletypesize{\scriptsize}
\tablecolumns{12}
\tablewidth{0pc}
\tablecaption{Sample and Radio Properties
\label{tab:sample}
}
\tablehead{
\colhead{} & 
\colhead{} & 
\colhead{} & 
\colhead{FIRST} & 
\colhead{FIRST} & 
\colhead{NVSS} & 
\colhead{WENSS} & 
\colhead{Angular} & 
\colhead{} & 
\colhead{} & 
\colhead{} & 
\colhead{} \\ 
\colhead{Object} & 
\colhead{} & 
\colhead{log$(5100\,$\AA$\,L_{5100})$} & 
\colhead{Core} & 
\colhead{Total} & 
\colhead{Total} & 
\colhead{Total\tablenotemark{a}} & 
\colhead{Size} & 
\colhead{log($L_{c}$)} & 
\colhead{log($L_{325}$)} & 
\colhead{} & 
\colhead{} \\ 
\colhead{SDSS J} & 
\colhead{Redshift} & 
\colhead{ergs s$^{-1}$ Hz$^{-1}$} & 
\colhead{mJy} & 
\colhead{mJy} & 
\colhead{mJy} & 
\colhead{mJy} & 
\colhead{\arcsec} & 
\colhead{ergs s$^{-1}$ Hz$^{-1}$} & 
\colhead{ergs s$^{-1}$ Hz$^{-1}$} & 
\colhead{log($R_{V}$)} & 
\colhead{$\alpha_{r}$} 
}
\startdata
073422.19+472918.8
&
0.3819
 & 
44.701$\pm$0.001\phantom{$\,<$}
 & 
\phantom{000}6.47
 & 
\phantom{0}213.8$\pm$0.14
 & 
\phantom{0}254.9$\pm$5.8\phantom{00}
 & 
\phantom{00}766
 & 
\phantom{00}90
 & 
31.37$\pm\!<$0.01
 & 
33.58$\pm\!<$0.01
 & 
\phantom{$<$}1.44$\pm$0.01\phantom{$\,<$}
 & 
$-$0.75$\pm$0.02\phantom{0}
 \\ 
074125.22+333319.9
&
0.3641
 & 
44.833$\pm\!<$0.001
 & 
\phantom{000}3.01
 & 
\phantom{0}391.7$\pm$0.22
 & 
\phantom{0}586.6$\pm$13.0\phantom{0}
 & 
\phantom{0}1853
 & 
\phantom{0}160
 & 
30.99$\pm$0.03\phantom{$\,<$}
 & 
33.92$\pm$0.01
 & 
\phantom{$<$}0.93$\pm$0.03\phantom{$\,<$}
 & 
$-$0.79$\pm$0.02\phantom{0}
 \\ 
074541.66+314256.6
&
0.4609
 & 
45.814$\pm\!<$0.001
 & 
\phantom{0}614.64
 & 
 1270.8$\pm$0.15
 & 
 1357.8$\pm$41.5\phantom{0}
 & 
\phantom{0}3458
 & 
\phantom{000}8
 & 
33.52$\pm\!<$0.01
 & 
34.43$\pm$0.01
 & 
\phantom{$<$}2.47$\pm\!<$0.01
 & 
$-$0.64$\pm$0.02\phantom{0}
 \\ 
075145.14+411535.8
&
0.4290
 & 
44.449$\pm$0.003\phantom{$\,<$}
 & 
\phantom{0}202.68
 & 
\phantom{0}224.6$\pm$0.14
 & 
\phantom{0}244.9$\pm$7.4\phantom{00}
 & 
\phantom{00}460
 & 
\phantom{0}310
 & 
32.97$\pm\!<$0.01
 & 
33.48$\pm\!<$0.01
 & 
\phantom{$<$}3.29$\pm$0.02\phantom{$\,<$}
 & 
$-$0.43$\pm$0.02\phantom{0}
 \\ 
080413.87+470442.8
&
0.5095
 & 
44.284$\pm$0.006\phantom{$\,<$}
 & 
\phantom{0}847.18
 & 
\phantom{0}869.6$\pm$0.21
 & 
\phantom{0}888.2$\pm$26.6\phantom{0}
 & 
\phantom{0}2574
 & 
\phantom{00}10
 & 
33.75$\pm\!<$0.01
 & 
34.41$\pm$0.01
 & 
\phantom{$<$}4.23$\pm$0.06\phantom{$\,<$}
 & 
$-$0.73$\pm$0.02\phantom{0}
 \\ 
080644.42+484149.2
&
0.3701
 & 
44.913$\pm\!<$0.001
 & 
\phantom{00}43.32
 & 
\phantom{0}840.5$\pm$0.14
 & 
\phantom{0}901.7$\pm$23.2\phantom{0}
 & 
\phantom{0}3099
 & 
\phantom{00}100
 & 
32.17$\pm\!<$0.01
 & 
34.16$\pm$0.01
 & 
\phantom{$<$}2.02$\pm\!<$0.01
 & 
$-$0.85$\pm$0.02\phantom{0}
 \\ 
080754.50+494627.6
&
0.5752
 & 
44.494$\pm$0.004\phantom{$\,<$}
 & 
\phantom{000}1.05
 & 
\phantom{0}298.2$\pm$0.21
 & 
\phantom{0}384.1$\pm$12.0\phantom{0}
 & 
\phantom{0}1452
 & 
\phantom{0}160
 & 
30.95$\pm$0.09\phantom{$\,<$}
 & 
34.29$\pm\!<$0.01
 & 
$<$1.2\phantom{$\pm$0.1\phantom{$\,<$}}
 & 
$-$0.91$\pm$0.02\phantom{0}
 \\ 
080814.70+475244.7
&
0.5455
 & 
45.153$\pm\!<$0.001
 & 
\phantom{00}26.39
 & 
\phantom{00}50.3$\pm$0.15
 & 
\phantom{00}69.3$\pm$2.8\phantom{00}
 & 
\phantom{00}226
 & 
\phantom{000}5
 & 
32.30$\pm\!<$0.01
 & 
33.43$\pm\!<$0.01
 & 
\phantom{$<$}1.92$\pm\!<$0.01
 & 
$-$0.81$\pm$0.03\phantom{0}
 \\ 
080833.36+424836.3
&
0.5429
 & 
44.463$\pm$0.004\phantom{$\,<$}
 & 
\phantom{00}27.35
 & 
\phantom{00}72.6$\pm$0.14
 & 
\phantom{00}91.4$\pm$3.1\phantom{00}
 & 
\phantom{00}279
 & 
\phantom{00}30
 & 
32.31$\pm\!<$0.01
 & 
33.51$\pm\!<$0.01
 & 
$<$2.62\phantom{$\pm$0.04\phantom{$\,<$}}
 & 
$-$0.76$\pm$0.02\phantom{0}
 \\ 
081058.99+413402.7
&
0.5067
 & 
44.659$\pm$0.002\phantom{$\,<$}
 & 
\phantom{0}169.98
 & 
\phantom{0}189.3$\pm$0.14
 & 
\phantom{0}219.7$\pm$6.6\phantom{00}
 & 
\phantom{00}390
 & 
\phantom{000}9
 & 
33.05$\pm\!<$0.01
 & 
33.58$\pm\!<$0.01
 & 
\phantom{$<$}3.16$\pm$0.02\phantom{$\,<$}
 & 
$-$0.39$\pm$0.02\phantom{0}
 \\ 
\enddata
\tablecomments{Table \ref{tab:sample} is published in its entirety in the electronic edition of ApJ, A portion is shown here for guidance regarding its form and content.  The uncertainty on the WENSS fluxes is taken to be 3.6~mJy, the 1$\sigma$ noise in the radio maps.}
\end{deluxetable}

%% file: spec_prop_short.tex
\begin{deluxetable}{lccc}
\tabletypesize{\scriptsize}
\tablecolumns{4}
\tablewidth{0pc}
\tablecaption{Spectral Properties
\label{tab:spec}
}
\tablehead{
\colhead{Object} & 
\colhead{FWHM \Hb} & 
\colhead{EW \Hb\tablenotemark{a}} & 
\colhead{EW \FeII\tablenotemark{a}} \\
\colhead{SDSS J} & 
\colhead{km s$^{-1}$} & 
\colhead{\AA} & 
\colhead{\AA} 
}
\startdata
073422.19+472918.8
	& 
\phantom{0}8700$\pm$400\phantom{00}
 & 
\phantom{00}142$\pm$7\phantom{0000}
 & 
\phantom{000}34$\pm$4\phantom{0000}
 \\ 
074125.22+333319.9
	& 
\phantom{0}8300$\pm$200\phantom{00}
 & 
\phantom{000}89$\pm$4\phantom{0000}
 & 
\phantom{0000}1$\pm$4\phantom{0000}
 \\ 
074541.66+314256.6
	& 
       10730$\pm$80\phantom{000}
 & 
\phantom{00}133$\pm$4\phantom{0000}
 & 
\phantom{000}12$\pm$3\phantom{0000}
 \\ 
075145.14+411535.8
	& 
\phantom{0}3000$\pm$2000\phantom{0}
 & 
\phantom{000}12$\pm$3\phantom{0000}
 & 
\nodata
 \\ 
080644.42+484149.2
	& 
       15200$\pm$700\phantom{00}
 & 
\phantom{000}43$\pm$3\phantom{0000}
 & 
\phantom{0000}4$\pm$4\phantom{0000}
 \\ 
080754.50+494627.6
	& 
       16000$\pm$2000\phantom{0}
 & 
\phantom{00}180$\pm$40\phantom{000}
 & 
\phantom{000}50$\pm$5\phantom{0000}
 \\ 
080814.70+475244.7
	& 
\phantom{0}9100$\pm$500\phantom{00}
 & 
\phantom{00}107$\pm$7\phantom{0000}
 & 
\phantom{000}32$\pm$4\phantom{0000}
 \\ 
080833.36+424836.3
	& 
\phantom{0}9200$\pm$800\phantom{00}
 & 
\phantom{00}150$\pm$10\phantom{000}
 & 
\phantom{000}63$\pm$4\phantom{0000}
 \\ 
081253.10+401859.9
	& 
\phantom{0}4000$\pm$2000\phantom{0}
 & 
\phantom{000}47$\pm$8\phantom{0000}
 & 
\phantom{000}27$\pm$5\phantom{0000}
 \\ 
081318.85+501239.7
	& 
\phantom{0}3400$\pm$400\phantom{00}
 & 
\phantom{00}170$\pm$10\phantom{000}
 & 
\phantom{000}49$\pm$4\phantom{0000}
 \\ 
\enddata
\tablecomments{Table \ref{tab:spec} is published in its entirety in the electronic edition of ApJ.  A portion is shown here for guidance regarding its form and content.  Equivalent widths are given in the rest frame.}
\end{deluxetable}

%% file: wb86_sample_short.tex
\begin{deluxetable}{lcccccc}
\tabletypesize{\scriptsize}
\tablecolumns{7}
\tablewidth{0pc}
\tablecaption{WB86 Sample Properties
\label{tab:sample_wb}
}
\tablehead{
\colhead{} & 
\colhead{RA} & 
\colhead{DEC} & 
\colhead{} & 
\colhead{FWHM} & 
\colhead{log$(L_{325})$} & 
\colhead{} \\ 
\colhead{Object} & 
\colhead{J2000} & 
\colhead{J2000} & 
\colhead{Redshift} & 
\colhead{km s$^{-1}$} & 
\colhead{erg s$^{-1}$} & 
\colhead{log$(R_{V})$}
}
\startdata
0003+158
 & 
 00 05 59.2
 & 
+16 09 49.0
 & 
0.4500
 & 
5680
 & 
34.36$\pm\!<$0.01
 & 
2.16
 \\ 
0007+106
 & 
 00 10 31.0
 & 
+10 58 29.5
 & 
0.0890
 & 
4300
 & 
30.7$\pm$0.5\phantom{$\,<$}
 & 
3.69
 \\ 
0044+030
 & 
 00 47 05.9
 & 
+03 19 54.9
 & 
0.6240
 & 
7700
 & 
34.12$\pm$0.02\phantom{$\,<$}
 & 
2.11
 \\ 
0106+380
 & 
 01 09 25.5
 & 
+38 16 45.0
 & 
0.5830
 & 
5300
 & 
33.88$\pm$0.02\phantom{$\,<$}
 & 
2.16
 \\ 
0109+200
 & 
 01 12 10.2
 & 
+20 20 21.8
 & 
0.7460
 & 
6140
 & 
34.54$\pm$0.02\phantom{$\,<$}
 & 
2.18
 \\ 
0133+207
 & 
 01 36 24.4
 & 
+20 57 27.4
 & 
0.4250
 & 
3200
 & 
34.97$\pm\!<$0.01
 & 
1.24
 \\ 
0134+329
 & 
 01 37 41.3
 & 
+33 09 35.1
 & 
0.3670
 & 
2700
 & 
35.34$\pm\!<$0.01
 & 
1.50
 \\ 
0139+391
 & 
 01 41 57.8
 & 
+39 23 29.1
 & 
0.0800
 & 
12500
 & 
31.6$\pm$0.1\phantom{$\,<$}
 & 
3.63
 \\ 
0145+210
 & 
 01 47 53.8
 & 
+21 15 39.7
 & 
0.6240
 & 
2870
 & 
34.11$\pm$0.05\phantom{$\,<$}
 & 
2.95
 \\ 
0159$-$117
 & 
 02 01 57.2
 & 
$-$11 32 33.2
 & 
0.6700
 & 
6150
 & 
35.12$\pm$0.01\phantom{$\,<$}
 & 
2.60
 \\ 
\enddata
\tablecomments{Table \ref{tab:sample_wb} is published in its entirety in the electronic edition of ApJ, A portion is shown here for guidance regarding its form and content.}
\end{deluxetable}

%% file: sed_sample_short.tex
\begin{deluxetable}{lcccccc}
\tabletypesize{\scriptsize}
\tablecolumns{7}
\tablewidth{0pc}
\tablecaption{SED Sample Properties
\label{tab:sample_sed}
}
\tablehead{
\colhead{} & 
\colhead{RA} & 
\colhead{DEC} & 
\colhead{} & 
\colhead{FWHM} & 
\colhead{log$(L_{325})$} & 
\colhead{} \\ 
\colhead{Object} & 
\colhead{J2000} & 
\colhead{J2000} & 
\colhead{Redshift} & 
\colhead{km s$^{-1}$} & 
\colhead{erg s$^{-1}$} & 
\colhead{log$(R_{V})$}
}
\startdata
3C110    
 & 
 04 17 16.7
 & 
$-$05 53 45.0
 & 
0.7749
 & 
12450
 & 
35.10$\pm$0.01\phantom{$\,<$}
 & 
2.24
 \\ 
3C175    
 & 
 07 13 02.4
 & 
+11 46 14.7
 & 
0.7693
 & 
20930
 & 
35.50$\pm\!<$0.01
 & 
1.15
 \\ 
3C186    
 & 
 07 44 17.5
 & 
+37 53 17.1
 & 
1.0630
 & 
\nodata
 & 
35.66$\pm$0.02\phantom{$\,<$}
 & 
3.06
 \\ 
3C207    
 & 
 08 40 47.6
 & 
+13 12 23.6
 & 
0.6797
 & 
3505
 & 
35.31$\pm$0.02\phantom{$\,<$}
 & 
3.09
 \\ 
3C215    
 & 
 09 06 31.9
 & 
+16 46 11.4
 & 
0.4108
 & 
6760
 & 
34.53$\pm\!<$0.01
 & 
1.83
 \\ 
3C232    
 & 
 09 58 21.0
 & 
+32 24 02.2
 & 
0.5297
 & 
4655
 & 
34.63$\pm$0.02\phantom{$\,<$}
 & 
2.78
 \\ 
3C254    
 & 
 11 14 38.7
 & 
+40 37 20.3
 & 
0.7363
 & 
14100
 & 
35.54$\pm$0.02\phantom{$\,<$}
 & 
1.98
 \\ 
3C263    
 & 
 11 39 57.0
 & 
+65 47 49.4
 & 
0.6464
 & 
4970
 & 
35.28$\pm$0.02\phantom{$\,<$}
 & 
2.04
 \\ 
3C277.1  
 & 
 12 52 26.4
 & 
+56 34 19.7
 & 
0.3199
 & 
3835
 & 
34.35$\pm$0.02\phantom{$\,<$}
 & 
2.08
 \\ 
3C281    
 & 
 13 07 54.0
 & 
+06 42 14.3
 & 
0.6017
 & 
7985
 & 
34.77$\pm$0.01\phantom{$\,<$}
 & 
2.21
 \\ 
\enddata
\tablecomments{Table \ref{tab:sample_sed} is published in its entirety in the electronic edition of ApJ, A portion is shown here for guidance regarding its form and content.}
\end{deluxetable}

%% file: b15_sample_short.tex
\begin{deluxetable}{lcccccc}
\tabletypesize{\scriptsize}
\tablecolumns{7}
\tablewidth{0pc}
\tablecaption{B15 Sample Properties
\label{tab:sample_b15}
}
\tablehead{
\colhead{Object} & 
\colhead{RA} & 
\colhead{DEC} & 
\colhead{} & 
\colhead{FWHM} & 
\colhead{log$(L_{325})$} & 
\colhead{} \\ 
\colhead{SDSS J} & 
\colhead{J2000} & 
\colhead{J2000} & 
\colhead{Redshift} & 
\colhead{km s$^{-1}$} & 
\colhead{erg s$^{-1}$} & 
\colhead{log$(R_{V})$}
}
\startdata
000111.19$-$002011.5
 & 
 00 01 11.2
 & 
$-$00 20 11.5
 & 
0.5179
 & 
5016
 & 
\nodata
 & 
2.37
 \\ 
005905.51+000651.6
 & 
 00 59 05.5
 & 
+00 06 51.6
 & 
0.7189
 & 
4976
 & 
\nodata
 & 
3.84
 \\ 
005917.47$-$091953.7
 & 
 00 59 17.5
 & 
$-$09 19 53.7
 & 
0.6409
 & 
3654
 & 
\nodata
 & 
2.53
 \\ 
010644.15$-$103410.5
 & 
 01 06 44.2
 & 
$-$10 34 10.5
 & 
0.4677
 & 
3355
 & 
\nodata
 & 
3.04
 \\ 
012905.32$-$005450.5
 & 
 01 29 05.3
 & 
$-$00 54 50.5
 & 
0.7067
 & 
3020
 & 
\nodata
 & 
1.81
 \\ 
013352.66+011345.1
 & 
 01 33 52.7
 & 
+01 13 45.1
 & 
0.3081
 & 
4205
 & 
\nodata
 & 
1.90
 \\ 
021125.07$-$081440.3
 & 
 02 11 25.1
 & 
$-$08 14 40.3
 & 
0.5371
 & 
4673
 & 
\nodata
 & 
2.51
 \\ 
021225.56+010056.1
 & 
 02 12 25.6
 & 
+01 00 56.1
 & 
0.5128
 & 
4945
 & 
\nodata
 & 
2.50
 \\ 
030210.95$-$075209.4
 & 
 03 02 10.9
 & 
$-$07 52 09.4
 & 
0.7338
 & 
5381
 & 
\nodata
 & 
3.61
 \\ 
073320.83+390505.1
 & 
 07 33 20.8
 & 
+39 05 05.1
 & 
0.6637
 & 
2759
 & 
33.72$\pm\!<$0.01
 & 
2.86
 \\ 
\enddata
\tablecomments{Table \ref{tab:sample_b15} is published in its entirety in the electronic edition of ApJ, A portion is shown here for guidance regarding its form and content.}
\end{deluxetable}

%% file: ks_results.tex
\begin{deluxetable}{ccccc}
\setlength{\tabcolsep}{12pt}
\tablecolumns{5}
\tablewidth{0pc}
\tablecaption{Sample Comparison Statistics
\label{tab:ks}
}
\tablehead{
\colhead{Sample} & 
\colhead{$\rho_{SR}$} &
\colhead{P$_{SR}$} &
\colhead{D$_{KS}$} & 
\colhead{P$_{KS}$} 
}
\startdata
This Work 	&$-$0.01 	& 0.90			& 0.06 & 0.76 				\\
WB86 		&0.13		& 0.26 			& 0.14 & 0.07 				\\
SED 		&0.26 		& 0.06			& 0.10 & 0.67 			   	\\
B15 		&0.26		& 2.0$\times10^{-7}$	& 0.13 & 2.2$\times10^{-6}$ 	\\
\enddata
\tablecomments{The Spearman-Rank correlation coefficient and associated probability for log$(R_{V})$ with redshift.  A probability of $<\!0.05$ indicates a statistically significant correlation.  One-sample KS test compares to the expected distribution of log$(R_{V})$ for sources distributed uniformly in solid angle.  A probability of $<\!0.05$ indicates that the sample is not drawn from the expected distribution.}
\end{deluxetable}